\documentclass[twocolumn]{article}

\usepackage{graphicx}
\usepackage{dcolumn}
\usepackage{bm}
\usepackage{amsmath}
\usepackage{amssymb}

\usepackage[utf8]{inputenc}
\usepackage[T1]{fontenc}
\usepackage{mathptmx}
\usepackage{authblk}

\usepackage[outdir=./Figures/]{epstopdf}
\usepackage[nolist]{acronym}
\usepackage{siunitx}
\usepackage{tabularx}
\usepackage{float}
\usepackage[labelformat=simple]{subcaption}

\graphicspath{{Figures/}}
\usepackage[para]{threeparttable}
\usepackage{placeins}
\usepackage{breqn}
\usepackage{xcolor}
\usepackage{hyperref}
\usepackage{cuted}

\usepackage[sort&compress,numbers]{natbib}
\usepackage{doi}

 \hypersetup{
      breaklinks=true,
      colorlinks=true,
       linkcolor=BrickRed,                   
       citecolor=black,                       
      linkcolor=black,                      
      filecolor=black,
      urlcolor=black
  }%
\usepackage{soul}

\makeatletter
\preto\maketitle{%
  \begingroup\lccode`~=`,
  \lowercase{\endgroup
  \let\saved@breqn@active@comma~
  \let~}\active@comma 
}
\appto\maketitle{%
  \begingroup\lccode`~=`,
  \lowercase{\endgroup
  \let~}\saved@breqn@active@comma 
}
\makeatother

\date{\small\today}
\author[1]{Duncan Dockar}
\author[2]{M. H. Lakshminarayana Reddy}
\author[1]{Matthew K. Borg}
\author[2]{S. Kokou Dadzie \thanks{k.dadzie@hw.ac.uk}}

\affil[1]{Institute for Multiscale Thermofluids, School of Engineering, University of Edinburgh, Edinburgh EH9 3FB, UK}
\affil[2]{School of Engineering and Physical Sciences, Heriot-Watt University, Edinburgh EH14 4AS, Scotland, UK}

\title{Volume diffusion modelling of a sheared granular gas}

\begin{acronym}
	\newacro{MD}{Molecular Dynamics}
	\newacro{DEM}{Discrete Element Method}
	\newacro{NEMD}{Non-Equilibrium Molecular Dynamics}
	\newacro{CFD}{Computational Fluid Dynamics}
	\newacro{LJ}{Lennard-Jones}
	\newacro{AFM}{Atomic Force Microscopy}
	\newacro{TPCL}{Three-Phase Contact Line}
	\newacro{CL}{Contact Line}
	\newacro{So}[S\textsubscript{o}]{hydrophobic}
	\newacro{Si}[S\textsubscript{i}]{hydrophilic}
	\newacro{FCC}{Face Centred Cubic}
	\newacro{CCA}{Constant Contact Angle}
	\newacro{CCR}{Constant Contact Radius}
	\newacro{H2O}[H\textsubscript{2}O]{water}
	\newacro{N2}[N\textsubscript{2}]{nitrogen}
	\newacro{PDMS}{polydimethylsiloxane}
	\newacro{2D}{two-dimensional}
	\newacro{3D}{three-dimensional}
	\newacro{VDW}{Redlich--Kwong}
	\newacro{VDW}{Van der Waals}
	\newacro{VD}{Volume Diffusion}
	\newacro{NSF}{Navier--Stokes--Fourier}
	\newacro{LE}{Lees--Edwards}
	\newacro{CS}{Carnahan--Starling}
\end{acronym}

\newcommand\Knn{\mbox{Kn}}  

\def\onedot{$\mathsurround0pt\ldotp$}
\def\cddot{
  \mathbin{\vcenter{\baselineskip.67ex
    \hbox{\onedot}\hbox{\onedot}}%
  }}%

\begin{document}
\maketitle
\begin{strip}
\vspace{-15mm}
\begin{abstract}
Continuum fluid dynamic models based on the \ac{NSF} equations have previously been used to simulate granular media undergoing fluid-like shearing. These models, however, typically fail to predict the flow behaviour in confined environments as non-equilibrium particle effects dominate near walls. We adapt an extended hydrodynamic model for granular flows, which uses a density-gradient dependent ``volume diffusion'' term to correct the viscous stress tensor and heat flux, to simulate the shearing of a granular gas between two rough walls, and with corresponding boundary conditions. We use our \ac{VD} model to predict channel flows for a range of mean volume fraction $\bar{\phi}=0.01$--$0.4$, and inter-particle coefficients of restitution $e=0.8$ and $0.9$, and compare with \ac{DEM} simulations and classical \ac{NSF} equations. Our model is capable of predicting non-uniform pressure, volume fraction and granular temperature, which become more significant for cases with mean volume fraction $\bar{\phi}\sim0.1$, in which we typically observe non-uniform peak density variations, and large volume fraction gradients.
\acresetall
\end{abstract}
\end{strip}

\section{Introduction}\label{sec:Intro}
Granular flows are frequently encountered in many environmental and industrial processes, such as: avalanches and landslides \cite{FriedmannRockAvalanches2006,CuomoModellingFlowslidesReview2020}, powder bed fusion \cite{SoundararajanPodwerBedFusionReviewModelling2021,HaeriDEMSpreading2017,HaeriOptimisationBladeSpreaders2017}, and grain transport and storage \cite{GanesanFlowabilityBulkSolidPowdersReview2008,JianSegregationBulkGrains2019}. Despite being a commonly encountered phenomenon, accurate modelling of granular flows is still a challenging problem due to their complex rheological properties: at low shear rates, particles settle into an ordered solid-like structures, while at higher shear rates, they can flow like a fluid \cite{BrennenGranularMultiphaseFlows2005,ForterreOlivierDenseGranularFlows2008,RaoGranularFlowIntro2008}. Researchers have identified how liquid-like properties can be observed for shearing high volume fraction flows, i.e., when the ratio of mean flow density to the solid particle density, $\bar{\phi}$, is between $0.4$ and $0.6$. The flows can then be modelled with continuum fluid methods \cite{ForterreOlivierDenseGranularFlows2008,CuomoModellingFlowslidesReview2020,BrennenGranularMultiphaseFlows2005}.
\par
As volume fraction, $\bar{\phi}$, decreases ($\lessapprox0.4$), particles undergo binary collisions \cite{daCruzMacroscopicFrictionGranular2003,ForterreOlivierDenseGranularFlows2008,BrennenGranularMultiphaseFlows2005}, and the granular flow exhibits more gas-like properties. \citet{SG1998} and \citet{GarzoDuftyGranularKinetic1999} investigated the fluid properties of a granular gas using kinetic theory, assuming inelastic particle collisions controlled by the coefficient of restitution, $e$, and determined an equivalent set of transport coefficients, including: shear viscosity, bulk viscosity, and thermal conductivity. Granular gas fluid properties differ from an ideal gas due to inelastic particle collisions as well as rotational inertia and particle roughness and size, all of which introduce energy dissipation modes not admitted in ideal gases \cite{GarzoDuftyGranularKinetic1999,SantosTraansportCoefficientsInelasticMaxwell2003,NottBCWallGranular2011}. Many studies have exploited gas-like behaviours to model the one-dimensional shearing of granular fluids undergoing inelastic collisions for smooth particles \cite{LunKineticGranularCouette1984}, rough particles \cite{AbuZaidKineticGranularFrictional1990,LunKineticGranularRoughInelastic1987,ChialvoKineticFrictionGranular2013}, and flows confined between two parallel walls \cite{VescoviGranularShearCouetteDEM2014,CampbellBrennenSimulationGranular1985}.
\citet{VescoviGranularShearCouetteDEM2014} employed continuum \acl{NSF}-like equations for granular flow modelling in a sheared channel with $\bar{\phi}\gtrsim0.1$, however, they needed to fit the slip velocity using data from their \ac{DEM} simulations in the more inelastic cases $(e>0.7)$, instead of the initially proposed boundary conditions derived by \citet{RichmanBC1988} for rough walls, in order to obtain good agreement with volume fraction and granular temperature variations across the channel, highlighting the difficulties of continuum granular flow modelling near boundaries.
\par
As the volume fraction decreases further, $\bar{\phi}\lessapprox0.1$, we enter the regime of non-equilibrium flows in confined spaces, which cannot be accurately modelled by existing continuum theories \cite{DadzieThermomechanicalContinuumFlow2013,WuGranularGasPoiseuille2016,DadzieBivelocityGasLidCavity2016,ChristouHeatTransferCavity2017,ChristouRarefiedGasRecasted2018,StamatiouEnhancedFlowNanotube2019,LockerbyBurnettMicroCouette2003}. Particle-based methods may be used to model such flows, however, are computationally expensive with less wide-spread adoption in industries. In elastic gases, non-equilibrium effects of rarefaction nature are characterised by the Knudsen number, $\Knn$, which is the ratio of particle mean free path to a characteristic length, e.g.,\ the height of a micro-channel.
Some noticeable effects of rarefaction typically occur in Knudsen layers (regions close to boundaries with thickness of order of the gas' mean free path) where predictions of local density, velocity, and temperature variations fail \cite{LockerbyBurnettMicroCouette2003,DadzieBivelocityGasLidCavity2016,ChristouHeatTransferCavity2017,ChristouRarefiedGasRecasted2018,Reddy2020,DadzieThermomechanicalContinuumFlow2013}.
\par
A classical example of deviation from classical continuum flow model predictions is the phenomenon of enhanced mass flow rate of pressure-driven micro-channel gas flows in the now famous Knudsen-minimum experiment \cite{KnudsenMinimum1909}, arising from enhanced slip velocities at higher Knudsen numbers $\Knn>1$, while the conventional \ac{NSF} fluid equations erroneously predicts a monotonically decreasing mass flow-rate \cite{DadzieEnhancedMicroChannels2012,ChristouRarefiedGasRecasted2018}. A similar Knudsen minimum can be observed in granular channel flows by numerically solving the generalised Enskog equation for inelastic collisions, however, this phenomenon disappears as the ratio of the channel height to particle diameter decreases and approaches unity, until the mass flow rate appears to be a solely monotonically increasing function of $\Knn$ \cite{WuGranularGasPoiseuille2016}. Following from this example, continuum-based models are expected to poorly capture non-equilibrium behaviours dominating granular gas flows even in straightforward configurations.
\par
To overcome the inadequacies of the \ac{NSF} equations (which strictly is only valid for flows in thermodynamic equilibrium), the \ac{VD} equations have been proposed, following Brenner's bi-velocity theory \cite{BrennerBivelocity2012}, which introduce an additional \textit{volume diffusion flux} term to account for the fact that compressible fluid flows achieve mechanical equilibrium much quicker than reaching thermodynamic equilibrium \cite{BrennerFluidMechanicsRestVD2012}. This flux term, which is dependent on the local density gradient, captures additional stresses and work done to the fluid, and \ac{VD} hydrodynamics correct the stress tensor and heat flux which can more accurately capture localised non-equilibrium effects \cite{DadzieThermomechanicalContinuumFlow2013,LakshminarayanaMolecuarDiffusivityShockWave2021,DadzieBivelocityGasLidCavity2016,ChristouHeatTransferCavity2017,ChristouRarefiedGasRecasted2018}. \ac{VD} equations have been shown to provide better agreement over conventional \ac{NSF} equations when comparing with experiments and \ac{MD} simulations of confined and non-equilibrium fluid flows, in configurations such as the pressure driven flow in micro-channels  \cite{DadzieEnhancedMicroChannels2012,ChristouRarefiedGasRecasted2018}, lid-driven cavities \cite{DadzieBivelocityGasLidCavity2016,ChristouHeatTransferCavity2017}, shock-waves (with asymmetric density variations) \cite{LakshminarayanaMolecuarDiffusivityShockWave2021}, and enhanced liquid flow-rates in carbon nanotubes \cite{StamatiouEnhancedFlowNanotube2019}.
\par
In this work, we now adapt the \ac{VD} equations to model the shearing of a granular flow confined between two rough walls, benchmarking against \ac{DEM} simulations. We modify the existing \ac{VD} equations  to be suitable for modelling granular flows, using transport coefficients derived from kinetic theory \cite{GarzoDuftyGranularKinetic1999,LunKineticGranularCouette1984,VescoviGranularShearCouetteDEM2014}, and reduce the equations to a series of ordinary differential equations with boundary-value constraints. We extend our \ac{DEM} modelling into low density ranges, $\bar{\phi}<0.1$, which have not been extensively investigated for granular flows, and demonstrate how our new model better captures non-uniform pressure, density, and granular temperature variations between the walls, as well as the normal and shear stresses acting on the walls when compared to the conventional \ac{NSF} equation predictions.
\par
The remainder of this article is organised into the following sections: in \S\ref{sec:DEM}, we outline the channel geometry and our \ac{DEM} simulation methodology; in \S\ref{sec:VDmodel} we derive our shear flow model from the \ac{VD} equations; in \S\ref{sec:ResultsDiscussions} we show results of our \ac{DEM} simulations and \ac{VD} model, and compare with the classical \ac{NSF} equations.
Finally, in \S\ref{sec:Conclusions}, we summarise the findings of the article, and where this research could be further explored.

\section{\ac{DEM} simulations} \label{sec:DEM}
\begin{figure*}[!htb]
	\begin{center}
		\includegraphics{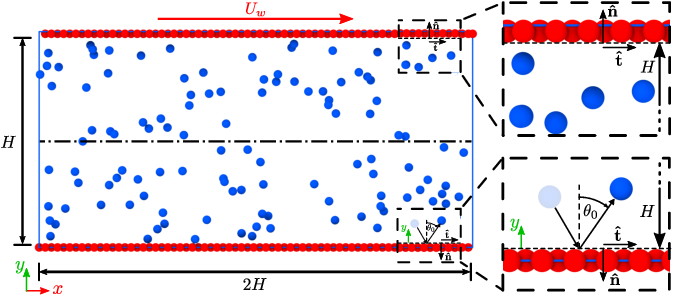}
		\caption{\ac{DEM} simulation set-up. Wall particles are shown in red, while the granular flow particles are shown in blue. The insets show the upper and lower rough boundary walls in more detail, and the lower inset also shows the definition of contact angle, $\theta_{0}$, used in Eqs.~\eqref{eq:NottSlip} and \eqref{eq:heatfluxBC0} for slip and heat flux boundary conditions, respectively. The dot-dashed line shows the centreline of the channel, at $y=H/2$.}
		\label{fig:DEMdomain}
	\end{center}
\end{figure*}

We simulated a granular flow shearing between two parallel particle walls using the \ac{DEM} software \textsc{LIGGGHTS} \cite{KlossLIGGGHTS2012}. We modelled a quasi-two-dimensional channel with height, $H=\SI{0.2475}{m}$, in the $y$ direction, and length $\SI{0.5}{m}(\approx2H)$, and $\SI{0.05}{m}$ thickness in the $x$ and $z$-directions, respectively, as shown in Fig.~\ref{fig:DEMdomain}.
The channel was enclosed between two single-layer \ac{FCC} walls, where the bottom wall was kept stationary and the top wall moved with constant velocity $U_{w}$.
\par
Periodic boundary conditions were used on the $x$ and $z$ domain faces, and the simulations were integrated through the velocity Verlet integration scheme \cite{KlossLIGGGHTS2012}, with a time-step $\Delta t=\SI{e-5}{s}$ \cite{daCruzMacroscopicFrictionGranular2003}. During the main production runs, granular particles were placed randomly within the channel walls to match a defined mean volume fraction, $\bar{\phi}$. The simulations were first run for $\SI{e9}{}$--$\SI{e10}{}\Delta t$, until they reached a steady-state in total system energy \cite{VescoviGranularShearCouetteDEM2014}. After this first run, the simulations were then run a further $\SI{5e7}{}\Delta t$, during which time we measured properties of the granular fluid locally across the channel width. This measurement period was chosen to be approximately $200\tau$, where $\tau=H/U_{w}$ is the characteristic shear-time, to ensure the observed \ac{DEM} simulation profiles were in a stable configuration \cite{daCruzMacroscopicFrictionGranular2003}.
Along with mean volume fraction $\bar{\phi}$, the coefficient of restitution was varied by investigating two values for inter-particle collisions $e=0.8$ and $0.9$. We chose not to model values below $e=0.8$, as this was found to lead to mostly asymmetric flow profiles, which is beyond the scope of this work, although could be investigated in the future using \ac{VD} modelling \cite{NottBoundariesCouetteBifurcation1999,AlamStabilityCouetteFlow1998,AlamInstabilityOrderingGravity2005}.
\par
In \ac{DEM} modelling, collision forces, $F_{i,j}$, between particles $i$ and $j$ are calculated by \cite{KlossLIGGGHTS2012}:
\begin{equation}
	\boldsymbol{F}_{i,j}=k_{n}\boldsymbol{r}_{i,j}-\gamma_{n}\dot{\boldsymbol{r}}_{i,j},
	\label{eq:DEMForce}
\end{equation}
where $\boldsymbol{r}_{i,j}$ and $\dot{\boldsymbol{r}}_{i,j}$ are the radial separation distance and relative normal velocity between particles $i$ and $j$, respectively. The normal stiffness, $k_{n}$, was set to $\SI{20}{kN/m}$, and the normal damping, $\gamma_{n}$, was calculated from the coefficient of restitution, $e$, as \cite{KlossLIGGGHTS2012}:
\begin{equation}
	\gamma_{n}=\sqrt{\frac{2k_{n}M}{1+\left(\frac{\pi}{\ln{e}}\right)^{2}}},
	\label{eq:RestitutionGamma}
\end{equation}
where $M=\rho_{0}\pi d_{0}^{3}/6$ is the mass of each particle, assuming equal particle masses. We specified particle diameter as, $d_{0}=\SI{0.01}{m}$, and solid grain density, $\rho_{0}=\SI{2500}{kg/m^{3}}$, and we neglected friction forces in collisions and tangentially acting stiffness and damping forces, which is equivalent to assuming perfectly smooth particles \cite{NottBCWallGranular2011}. For simplicity, the wall particles were specified to have the same properties as the interior granular fluid, i.e.\ identical diameter $d_{0}$, coefficient of restitution $e$, and grain density $\rho_{0}$.
\par
We measured macroscopic granular properties locally in ``bins'', such as the granular temperature:
\begin{equation}
	T=\frac{1}{3N}\left[\sum_{i=1}^{N}{\left|\left(\boldsymbol{u}_{i}-\boldsymbol{u}\right)\right |^{2}}\right],
	\label{eq:GranularTemperature}
\end{equation}
where $\boldsymbol{u}_{i}$ is velocity vector of a particle $i$ \cite{VescoviGranularShearCouetteDEM2014}, $N$ is the total number of particles in the bin, and the mean mass flow velocity in the bin is found by:
\begin{equation}
	\boldsymbol{u}=\frac{1}{N}{\sum_{i=1}^{n}{\boldsymbol{u}_{i}}}.
	\label{eq:Meanvelocity}
\end{equation}
Temperature is often measured in Kelvin, however, granular temperature in Eq.~\eqref{eq:GranularTemperature} has units $\SI{}{m^{2}/s^{2}}$; we will refer to the granular temperature simply as ``temperature'' from here on.
\par
The virial stress tensor $\mathbf{S}$ is obtained from the \ac{DEM} simulations in each bin:
\begin{equation}
	\mathbf{S}=\frac{1}{V}\left[\sum_{i}^{N}{M\left(\boldsymbol{u}_{i}-\boldsymbol{u}\right)\otimes\left(\boldsymbol{u}_{i}-\boldsymbol{u}\right)+\frac{1}{2}\sum_{i}^{N}{\sum_{j\neq i}^{N}{\boldsymbol{r}_{i,j}\otimes\boldsymbol{F}_{i,j}}}}\right],
	\label{eq:VirialStressTensor}
\end{equation}
where $V$ is the bin volume.\footnote{In \ac{DEM} simulations, the stress tensor is sometimes measured as the negative of the formulation given in Eq.~\eqref{eq:VirialStressTensor} \cite{KlossLIGGGHTS2012} Our definition of stress tensor in Eq.~\eqref{eq:VirialStressTensor} is chosen to be consistent with the stress tensor used in the \ac{NSF} and \ac{VD} equations, given later in \S\ref{sec:VDmodel}.}.
\par
In the \ac{DEM} simulations, the thermodynamic pressure $p$, which is used to relate the volume fraction and granular temperature, e.g.\ using the \ac{CS} radial distribution function \cite{VescoviGranularShearCouetteDEM2014,CarnahnStarlingg01969}, can be estimated from the trace of the virial stress tensor \cite{KlossLIGGGHTS2012}:
\begin{equation}
	p=\frac{1}{3}\left(S_{xx}+S_{yy}+S_{zz}\right),
	\label{eq:PDEM}
\end{equation}
where $S_{xx}$, $S_{yy}$, and $S_{zz}$ are the normal-face components of the stress tensor, in the $x$, $y$, and $z$ directions, respectively.

\section{Volume diffusion modelling for granular flows}\label{sec:VDmodel}
Our new hydrodynamic model for granular flows is adapted from the \acf{VD} model \cite{DadzieContinuumGasDensityVariation2008,DadzieThermomechanicalContinuumFlow2013}.
This adaptation of \ac{VD} equations for granular flows starts with the conservation of mass:
\begin{equation}
	\frac{\partial\rho}{\partial t}+\boldsymbol{\nabla}\cdot\left(\rho\boldsymbol{u}\right)=0,
	\label{eq:VDmass}
\end{equation}
where $\rho$ is the flow density, and $t$ is time. For a granular flow, density is more commonly expressed in terms of the grain density $\rho_{0}$, and the flow volume fraction $\phi$, i.e.\ $\rho=\rho_{0}\phi$ \cite{VescoviGranularShearCouetteDEM2014,BerziExtendedKineticTheory2014,JenkinsBerziKineticTheoryGranularInclined2012,BrennenGranularMultiphaseFlows2005}. \par
Next, we introduce the momentum transport equation \cite{DadzieContinuumGasDensityVariation2008,DadzieThermomechanicalContinuumFlow2013,ChristouRarefiedGasRecasted2018,DadzieBivelocityGasLidCavity2016,ChristouHeatTransferCavity2017,LakshminarayanaMolecuarDiffusivityShockWave2021}:
\begin{equation}
	\frac{\partial\left(\rho\boldsymbol{u}\right)}{\partial t}+\boldsymbol{\nabla}\cdot\left(\rho\boldsymbol{u}\otimes\boldsymbol{u}\right)+\boldsymbol{\nabla}\cdot\left[p\mathbf{I}+\boldsymbol{\Pi}\right]=\boldsymbol{0},
	\label{eq:VDmomentum}
\end{equation}
where $\mathbf{I}$ is the identity tensor, and $\boldsymbol{\Pi}$ is the accompanying stress tensor, as given by:
\begin{multline}
	\boldsymbol{\Pi}=-2\eta\left[\frac{1}{2}\left(\boldsymbol{\nabla}\left(\boldsymbol{u}-\boldsymbol{J}_{c}\right)+\left\{\boldsymbol{\nabla}\left(\boldsymbol{u}-\boldsymbol{J}_{c}\right)\right\}^{\top}\right)\right.\\
	\left.-\frac{1}{3}\mathbf{I}\boldsymbol{\nabla}\cdot\left(\boldsymbol{u}-\boldsymbol{J}_{c}\right)\right]
 	-\rho\boldsymbol{J}_{c}\otimes\boldsymbol{J}_{c}-\eta_{b}\mathbf{I}\boldsymbol{\nabla}\cdot\left(\boldsymbol{u}-\boldsymbol{J_{c}}\right),
	\label{eq:VDstresstensor}
\end{multline}
where $\eta$ is the shear viscosity and $\eta_{b}$ is the bulk viscosity, which have been derived by \citet{GarzoDuftyGranularKinetic1999} and listed in Appendix~\ref{Appendix:Coefficients}, along with other fluid transport coefficients.
The superscript $\top$ in Eq.~\eqref{eq:VDstresstensor} indicates the transpose of the given vector or tensor. Note that the sum of the pressure and \ac{VD} viscous stress tensor is equivalent to the virial stress tensor given in Eq.~\eqref{eq:VirialStressTensor}, for the \ac{DEM} simulations, i.e.\ $p\mathbf{I}+\boldsymbol{\Pi}=\mathbf{S}$.
\par
In Eq.~\eqref{eq:VDstresstensor} appears the contribution of the volume diffusion flux, $\boldsymbol{J}_{c}$, which arises from the kinetic level non-local  equilibrium  \cite{DadzieContinuumGasDensityVariation2008}.
This volume diffusion flux for the granular fluid may be expressed as:
\begin{equation}
	\boldsymbol{J}_{c}=-k_{m}\frac{\boldsymbol{\nabla}\rho}{\rho}=-k_{m}\frac{\boldsymbol{\nabla}\phi}{\phi},
	\label{eq:Jc}
\end{equation}
which we elaborate on more below.
The volume diffusivity coefficient, $k_{m}$, in Eq.~\eqref{eq:Jc} is expressed as \cite{ChristouRarefiedGasRecasted2018,DadzieBivelocityGasLidCavity2016,ChristouHeatTransferCavity2017}:
\begin{equation}
	k_{m}=\alpha^{*}\Pr\frac{\eta}{\rho_{0}\phi},
	\label{eq:kappamalpha}
 \end{equation}
where $\alpha^{*}$ is a dimensionless constant, and $\Pr=5\eta/2\kappa$ is the Prandtl number for a granular gas. Note that $\Pr$ varies from the typical formulation for an ideal gas because of the change in units when granular temperature and thermal conductivity, $\kappa$, are used (see Eq.~\eqref{eq:kineticheatfluxVec}). Previous investigations have found good agreement for gases using $\alpha^{*}=1/3$ \cite{ChristouRarefiedGasRecasted2018,DadzieBivelocityGasLidCavity2016}, however, for liquids and denser media, this coefficient decreases \cite{StamatiouEnhancedFlowNanotube2019,TomyDiffusionSlip2022}. In the present granular simulations, we found the approximation $\alpha^{*}=(3g_{0})^{-1}$ was sufficient, where $g_{0}$ is the radial distribution function  \cite{CarnahnStarlingg01969,VescoviGranularShearCouetteDEM2014}.
\par
\par
In \ac{VD} theory, the velocity $\boldsymbol{u}$ used here (sometimes referred to as the mass velocity) describes the motion of the mass of the constituting particles, as in the continuity equation in Eq.~\eqref{eq:VDmass}. However, this is not necessarily the same as the velocity of the deformable fluid element that contains the mass of a group of particles, which is instead referred to as the volume velocity $\boldsymbol{u}_{V}$ \cite{LakshminarayanaRNS2019,ChristouRarefiedGasRecasted2018,BrennerBivelocity2012,BrennerFluidMechanicsRestVD2012,DadzieThermomechanicalContinuumFlow2013}. The volume diffusion flux relates these two velocity terms by $\boldsymbol{J}_{c}=\boldsymbol{u}-\boldsymbol{u}_{V}$, and accounts for additional macroscopic motion of fluid that occurs, despite no net motion of the fluid mass.
\par
In its kinetic form, contributions from the volume diffusion flux to the constitutive equations are the direct consequence of molecular spatial stochasticity, not trivial in original kinetic equations such as the Boltzmann equation \cite{DadzieReeseThermomechanicalHydrodynamicKnudsen2012,DadzieReeseSpatialStochasityGasFlows2012}.
We may also note that these density gradient terms (i.e.\ Korteweg-type stress tensor \cite{LakshminarayanaRNS2019}) are the same ones obtained as quantum effects in the macroscopic continuum \acl{NSF} equations when the Madelung equations are derived from the Schr\"odinger equation \cite{BerloffFluidMechanicaNonZerpTemp2014}.
By setting $\boldsymbol{J}_{c}=0$ in Eq.~\eqref{eq:VDstresstensor}, we obtain the classical form of the stress tensor used in the classical \ac{NSF} granular model.
\par
Continuing with our transport equations, the energy equation for the granular gas is here proposed as the energy equation from \ac{VD} theory and complemented with energy dissipation from particle collisions  \cite{GarzoDuftyGranularKinetic1999,DadzieContinuumGasDensityVariation2008,DadzieEnhancedMicroChannels2012,DadzieReeseThermomechanicalHydrodynamicKnudsen2012,DadzieThermomechanicalContinuumFlow2013,ChristouRarefiedGasRecasted2018,DadzieBivelocityGasLidCavity2016,LakshminarayanaMolecuarDiffusivityShockWave2021}. That is:
\begin{multline}
		\frac{\partial}{\partial t}\left[\frac{1}{2}\rho u^{2}+\frac{3}{2}\rho T\right]+\boldsymbol{\nabla}\cdot\left[\frac{1}{2}\rho u^{2}\boldsymbol{u}+\frac{3}{2}\rho T\boldsymbol{u}\right]\\
		+\boldsymbol{\nabla}\cdot\left[\left(p\mathbf{I}+\boldsymbol{\Pi}\right)\cdot\boldsymbol{u}\right]+\boldsymbol{\nabla}\cdot\boldsymbol{J}_{u}+\frac{\Gamma}{L}=0,
	\label{eq:VDenergy}
\end{multline}
where $\Gamma$ is the rate of energy dissipation from particle collisions \cite{GarzoDuftyGranularKinetic1999}, and $L$ is the correlation length, a factor that accounts for increased particle collision frequency, when the assumption of molecular chaos breaks down, typically as local volume fraction increases, $\phi\gtrapprox0.4$ \cite{VescoviGranularShearCouetteDEM2014,BerziExtendedKineticTheory2014,JenkinsBerziKineticTheoryGranularInclined2012}.
We employed the functional expression \cite{BerziExtendedKineticTheory2014,JenkinsBerziKineticTheoryGranularInclined2012}:
\begin{equation}
	L=
	\begin{cases}
		1,& L^{*}\leq1\\
		L^{*},& L^{*}>1
	\end{cases}
\end{equation}
where:
\begin{equation}
	L^{*}=\frac{1}{2}(\phi g_{0})^{\frac{1}{3}}d_{0}\frac{\partial u}{\partial y}T^{-\frac{1}{2}}.
	\label{eq:CorrelationLength}
\end{equation}
The factor $L^{*}$ is only relevant for a few of our high volume fraction cases, although we include it to obtain good agreement, especially for the peak volume fraction variations expected in the channel flow \cite{VescoviGranularShearCouetteDEM2014,AlamStabilityCouetteFlow1998,NottBoundariesCouetteBifurcation1999,AlamInstabilityOrderingGravity2005}; our focus still remains on the non-equilibrium granular gas dynamics that become dominant for lower flow densities.
\par
Continuing from Eq.~\eqref{eq:VDenergy}, the energetic heat flux, $\boldsymbol{J}_{u}$, is given by \cite{DadzieContinuumGasDensityVariation2008,DadzieThermomechanicalContinuumFlow2013,ChristouHeatTransferCavity2017,DadzieBivelocityGasLidCavity2016,ChristouRarefiedGasRecasted2018}:
\begin{equation}
	\boldsymbol{J}_{u}=\boldsymbol{q}+p\boldsymbol{J}_{c},
	\label{eq:NewHeatFluxVec}
\end{equation}
where the classical heat flux (or the Fourier equivalent heat flux) for a granular gas is given by \cite{GarzoDuftyGranularKinetic1999}:
\begin{equation}
	\boldsymbol{q}=-\kappa\boldsymbol{\nabla}T-\mu\boldsymbol{\nabla}\phi,
	\label{eq:kineticheatfluxVec}
\end{equation}
and $\kappa$ is the thermal conductivity. Note that this thermal conductivity refers to the macroscopic properties of the granular gas and not to the solid grain material composing individual particles. Additionally, since we consider granular temperature in this work given in Eq.~\eqref{eq:GranularTemperature} with units $\SI{}{m^{2}/s^{2}}$, the equivalent SI units of $\kappa$ here are in $\SI{}{Ws^{2}/m^{3}}$. The second term in Eq.~\eqref{eq:kineticheatfluxVec} contains a coefficient $\mu$, that arises specifically in granular gases undergoing inelastic collisions, i.e.\ for $e<1$, and is not present in an ideal gas \cite{GarzoDuftyGranularKinetic1999,Reddy2015}.
\par
Eqs.~\eqref{eq:VDmass}, \eqref{eq:VDmomentum}, \eqref{eq:VDstresstensor}, and \eqref{eq:VDenergy} represent the \ac{VD} hydrodynamic equations which account for the additional volume diffusion flux, $J_{c}$, on the stress tensor and local heat flux \cite{DadzieBivelocityGasLidCavity2016,ChristouHeatTransferCavity2017,ChristouRarefiedGasRecasted2018,Reddy2020,DadzieThermomechanicalContinuumFlow2013,Reddy2020,LakshminarayanaRNS2019,LakshminarayanaMolecuarDiffusivityShockWave2021}. By writing them in this form, we can simply obtain the classical \ac{NSF} equations again, by setting $J_{c}=0$, which will become useful in the analysis of our \ac{DEM} and \ac{VD} results presented in \S\ref{sec:ResultsDiscussions}.
\par
For our sheared granular flow cases, as described in \S\ref{sec:DEM}, we assume a steady-state flow, i.e.\ $\partial/\partial t=0$, and with variations only in the $y$-direction, i.e.\ $\partial/\partial x, \partial/\partial z=0$. Only the $x$ component velocity $u$ is considered, so $\boldsymbol{u}=[u(y)\ 0\ 0]^{\top}$. Applying these assumptions, we write $\boldsymbol{J}_{c}=[0\ J_{c}(y)\ 0]^{\top}$ from Eq.~\eqref{eq:Jc}, where $J_{c}(y)$ is the $y$ component of $\boldsymbol{J}_{c}$. The stress tensor in Eq.~\eqref{eq:VDstresstensor} then becomes:
	\begin{multline}
		\boldsymbol{\Pi}=\\
		\footnotesize\left[\begin{matrix}
			 \left(-\frac{2\eta}{3}+\eta_{b}\right)\frac{\partial J_{c}}{\partial y} & -\eta\frac{\partial u}{\partial y} & 0 \\ -\eta\frac{\partial u}{\partial y} &
			 \left(\frac{4\eta}{3}+\eta_{b}\right)\frac{\partial J_{c}}{\partial y} -\rho_{0}\phi J_{c}^{2} & 0 \\
			 0 & 0 & \left(-\frac{2\eta}{3}+\eta_{b}\right)\frac{\partial J_{c}}{\partial y}
		\end{matrix}\right].
		\label{eq:VDstresstensorreduced}
	\end{multline}
By considering Eqs.~\eqref{eq:VDmomentum} and \eqref{eq:VDstresstensorreduced} in the $x$ direction, we denote the $xy$ shear stress component of the \ac{VD} virial stress tensor in Eq.~\eqref{eq:VDstresstensor} as $s=-S_{xy}$, which is constant, i.e.\ $\partial s/\partial y=0$, and:
\begin{equation}
	s=\eta\frac{\partial u}{\partial y}.
	\label{eq:stressviscosity}
\end{equation}
Similarly, in the $y$ direction, we extract:
\begin{equation}
	p+\left(\frac{4\eta}{3}+\eta_{b}\right)\frac{\partial J_{c}}{\partial y} -\rho_{0}\phi J_{c}^{2}=S_{yy},
	\label{eq:pressuregradient0}
\end{equation}
where $S_{yy}$ is a constant, i.e.\ $\partial S_{yy}/\partial y =0$. However, unlike in previous analyses \cite{VescoviGranularShearCouetteDEM2014,LunKineticGranularCouette1984,AbuZaidKineticGranularFrictional1990,NottBCWallGranular2011}, we find that the pressure $p$ is \textit{not} constant, owing to the volume diffusion flux, and is related to density and granular temperature by:
\begin{equation}
	p=fT,
	\label{eq:kineticpressure}
\end{equation}
where $f=\rho_{0}\phi[2g_{0}\phi(1+e)+1]$. We emphasise here, that it is important to distinguish between the normal stress $S_{yy}$, and the thermodynamic pressure $p$, since the normal stress is the physical force per unit area exerted on the particles within the channel and on the walls. The thermodynamic pressure accounts for the stress arising from the thermal or kinetic energy of the particles, from Eq.~\eqref{eq:kineticpressure}, however, as we will see in \S\ref{sec:ResultsDiscussions}, only partially contributes to the total normal stress, and instead we must consider the additional stress from the volume diffusion flux terms, see Eq.~\eqref{eq:pressuregradient0}, to accurately capture this normal stress profile. Still, we must carefully consider the thermodynamic pressure in our modelling, since it relates important thermodynamic properties together, such as volume fraction and temperature, particularly in high volume fraction cases, where the assumption of molecular chaos breaks down \cite{VescoviGranularShearCouetteDEM2014,BerziExtendedKineticTheory2014,JenkinsBerziKineticTheoryGranularInclined2012}.
\par
We employ the radial distribution function proposed by \citet{VescoviGranularShearCouetteDEM2014}, which is suitable for high volume fraction cases $\phi>0.4$, and highly inelastic collisions $e<0.95$:
\begin{equation}
	g_{0}(\phi)=\left\{\begin{matrix}
	g_{0_{CS}}(\phi), & \mathrm{ for\ } \phi\leq\phi_{m}
	\\
	\psi g_{0_{CS}}(\phi)+\frac{2\left(1-\psi\right)}{\phi_{rcp}-\phi}, & \mathrm{ for\ } \phi>\phi_{m}
	\end{matrix}\right.,
\label{eq:RDF}
\end{equation}
where $g_{0_{CS}}(\phi)=(2-\phi)/2(1-\phi)^{3}$ is the \ac{CS} radial distribution function \cite{CarnahnStarlingg01969}, and:
\begin{equation}
	\psi=\frac{\phi^{2}-2\phi_{m}\phi+\phi_{rcp}\left(2\phi_{m}-\phi_{rcp}\right)}{2\phi_{rcp}\phi_{m}-\phi_{m}^{2}-\phi_{rcp}^{2}},
	\label{eq:VescoviFactor}
\end{equation}
where $\phi_{rcp}=0.636$, $\phi_{f}=0.49$, and $\phi_{m}=0.4$, are the random close packing limit for hard spheres, volume fraction at ``freezing'' point, and volume fraction at ``merging'' point, respectively \cite{VescoviGranularShearCouetteDEM2014}.
In Appendix~\ref{Appendix:LeesEdwardsBC}, we verify that the radial distribution function given in Eq.~\eqref{eq:RDF} is suitable for our work, using data from \ac{DEM} simulations with imposed \ac{LE} boundary conditions \cite{LeesEdwardsBC1972,KlossLIGGGHTS2012}.
\par
Continuing our derivation, by reducing Eq.~\eqref{eq:VDenergy}, we obtain:
\begin{equation}
	s\frac{\partial u}{\partial y}=\frac{\partial q}{\partial y}+\frac{\partial}{\partial y}\left(pJ_{c}\right)+\frac{\Gamma}{L},
	\label{eq:VDenergy1D0}
\end{equation}
%
and heat flux is simplified from Eq.~\eqref{eq:kineticheatfluxVec}:
\begin{equation}
	q=-\kappa\frac{\partial T}{\partial y}-\mu\frac{\partial\phi}{\partial y},
	\label{eq:kineticheatflux}
\end{equation}
\par

Eq.~\eqref{eq:kineticpressure} is differentiated, and rearranged along with Eqs.~\eqref{eq:stressviscosity}--\eqref{eq:kineticpressure}, \eqref{eq:VDenergy1D0}, and \eqref{eq:kineticheatflux}. The full \ac{VD} hydrodynamics equations for our granular flow configuration, reduced into a set of differential equations, then become as follows:
\begin{gather}
	\frac{\partial u}{\partial y}=\frac{s}{\eta},
	\label{eq:VelocityDiff}\\
	\frac{\partial p}{\partial y}=\frac{J_{c}\phi}{k_{m}}\left(\frac{f\mu}{\kappa}-\frac{\partial f}{\partial\phi}\frac{p}{f}\right)-\frac{f}{\kappa}q,
	\label{eq:PressureDiff}\\
	\frac{\partial \phi}{\partial y}=-\frac{J_{c}\phi}{k_{m}},
	\label{eq:VoidDiff}\\
	\frac{\partial J_{c}}{\partial y}=\left(S_{yy}+\rho_{0}\phi J_{c}^{2}-p\right)\left(\frac{4\eta}{3}+\eta_{b}\right)^{-1},
	\label{eq:RecastedVelocityDiff}
\end{gather}
and:
\begin{equation}
	\frac{\partial q}{\partial y}=\frac{s^{2}}{\eta}-p\frac{\partial J_{c}}{\partial y}-J_{c}\frac{\partial p}{\partial y}-\frac{\Gamma}{L}.
	\label{eq:HeatFluxDiff}
\end{equation}
In order to compare with classical \ac{NSF} equations, we set $J_{c},\ \partial p/\partial y,\ \partial J_{c}/\partial y=0$, to obtain a set of alternative differential equations for \ac{NSF} modelling, and Eqs.~\eqref{eq:VoidDiff} and \eqref{eq:HeatFluxDiff} are instead replaced with:
\begin{equation}
	\frac{\partial \phi}{\partial y}=q\left(\frac{\kappa}{f}\frac{\partial f}{\partial\phi}T-\mu\right)^{-1},
	\label{eq:VoidDiffNS}
\end{equation}
and
\begin{equation}
	\frac{\partial q}{\partial y}=\frac{s^{2}}{\eta}-\frac{\Gamma}{L},
	\label{eq:HeatFluxDiffNS}
\end{equation}
respectively \cite{VescoviGranularShearCouetteDEM2014}, while Eq.~\eqref{eq:VelocityDiff} is used as before, and $p=S_{yy}$.
\par
To ensure both the \ac{VD} and \ac{NSF} models achieve a target mean density $\bar{\phi}$, which is chosen to match each case from our \ac{DEM} simulations, we introduce the term $m=\int_{0}^{y}\phi\mathrm{d}y$ \cite{VescoviGranularShearCouetteDEM2014}, such that:
\begin{equation}
	\frac{\partial m}{\partial y}=\phi.
	\label{eq:HoldupDiff}
\end{equation}
\par
Eqs.~\eqref{eq:VelocityDiff}--\eqref{eq:HoldupDiff} are now in the form of six explicit differential equations that can be solved. We must also obtain two additional parameters $s$ and $S_{yy}$, making a total of eight boundary conditions necessary to fully close the problem. We employ boundary conditions proposed by \citet{NottBCWallGranular2011} for smooth inelastic particles, to estimate slip velocity $u_{s}$ and wall heat flux $q_{w}$. The flux of linear momentum to the wall $\boldsymbol{\Lambda}_{w}$ is given by \cite{NottBCWallGranular2011}:
	\begin{multline}
		\boldsymbol{\Lambda}_{w}=\rho_{0}\phi g_{w}T\hat{\boldsymbol{n}}+\frac{2\sqrt{2}}{3\sqrt{\pi}}\rho_{0}\phi g_{w}T^{1/2}\mathbf{P}\cdot\boldsymbol{u_{s}}\\
		-\frac{1}{3}\rho_{0}d_{0}Tg_{w}\mathbf{P}\cdot\boldsymbol{\nabla}\phi\\
		-\rho_{0}d_{0}g_{w}\left[\frac{5}{96\sqrt{2}g_{0}}\left(1+\frac{12}{5}G\right)+\frac{\phi}{3}\right]\mathbf{P}\cdot\boldsymbol{\nabla}T\\
		-\rho_{0}d_{0}T^{1/2}g_{w}\left[\frac{5\sqrt{\pi}}{96g_{0}}\left(1+\frac{8}{5}G\right)\mathbf{Q}+\frac{2\sqrt{2}\phi}{\sqrt{\pi}}\mathbf{R}\right]\cddot\boldsymbol{\nabla}\boldsymbol{u},
		\label{eq:WallMomentumFlux}
	\end{multline}
where $\boldsymbol{u}_{s}=\boldsymbol{u}-\boldsymbol{U}_{w}$ is the slip velocity vector, $\boldsymbol{U}_{w}$ is the wall velocity vector, $G=\phi g_{0}$, and $g_{w}$ is a factor which quantifies the increase in volume fraction due to particles in contact with the wall, compared to that of the bulk \cite{NottBCWallGranular2011,RaoGranularFlowIntro2008}, and is given in Table~\ref{tab:Coefficients}. We have omitted rotational flux terms, which are only relevant for rough particles \cite{NottBCWallGranular2011}.
Coefficients $\mathbf{P}$, $\mathbf{Q}$, and $\mathbf{R}$, are given by:
\begin{subequations}
	\begin{equation}
		 	 \mathbf{P}=\frac{2\left(1-\cos^{3}{\theta_{0}}\right)}{\sin^{2}{\theta_{0}}}\hat{\boldsymbol{n}}\hat{\boldsymbol{n}}+\left[2\left(1+\cos{\theta_{0}}\right)^{-1}-\cos{\theta_{0}}\right]\hat{\boldsymbol{t}}\hat{\boldsymbol{t}},
	 	 \end{equation}
		\begin{multline}
			 \mathbf{Q}=\left(\frac{1}{3}+\cos^{2}{\theta_{0}}\right)\hat{\boldsymbol{n}}\hat{\boldsymbol{n}}\hat{\boldsymbol{n}}+\frac{1}{2}\sin^{2}{\theta_{0}}\left(\hat{\boldsymbol{t}}\hat{\boldsymbol{n}}\hat{\boldsymbol{t}}+\hat{\boldsymbol{t}}\hat{\boldsymbol{t}}\hat{\boldsymbol{n}}\right)\\
			 -\left(\frac{1}{6}+\frac{1}{2}\cos^{2}{\theta_{0}}\right)\hat{\boldsymbol{n}}\hat{\boldsymbol{t}}\hat{\boldsymbol{t}},
		 \end{multline}
		\begin{equation}
			 \mathbf{R}=\frac{1}{2}\left(1+\cos^{2}{\theta_{0}}\right)\hat{\boldsymbol{n}}\hat{\boldsymbol{n}}\hat{\boldsymbol{n}}+\frac{1}{4}\sin^{2}{\theta_{0}}\left(\hat{\boldsymbol{t}}\hat{\boldsymbol{n}}\hat{\boldsymbol{t}}+\hat{\boldsymbol{t}}\hat{\boldsymbol{t}}\hat{\boldsymbol{n}}+\hat{\boldsymbol{n}}\hat{\boldsymbol{t}}\hat{\boldsymbol{t}}\right),
	\end{equation}
	\label{eq:NottTensors}
\end{subequations}
respectively, where $\theta_{0}$ is the maximum angle that a particle can subtend normal to the wall, pertained by the wall's roughness \cite{NottBCWallGranular2011,RichmanBC1988}, and unit vectors $\hat{\boldsymbol{n}}$ and $\hat{\boldsymbol{t}}$, are the normal and tangential flow (following $x$) directions, respectively (see insets in Fig.~\ref{fig:DEMdomain}). Note vector $\hat{\boldsymbol{n}}$ points outwards from the wall, i.e.\ away from the channel. Unit vectors in the tangential $z$ direction have not been given in Eq.~\eqref{eq:NottTensors} to improve clarity, however, can be found in \citet{NottBCWallGranular2011}.
\par
\begin{figure}[!htb]
	\begin{center}
		\includegraphics[width=0.5\textwidth]{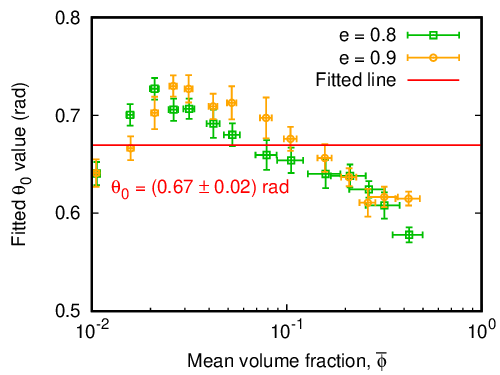}
		\caption{Variation in values of $\theta_{0}$, found by fitting Eq.~\eqref{eq:NottSlip} to our \ac{DEM} simulations, for different coefficients of restitution, $e$, and the solid line shows the mean value of all cases.}
		\label{fig:theta0fit}
	\end{center}	
\end{figure}

For our shear flow case, considering only the linear shear stress in the $x$ direction, i.e.\ $s$, we can rearrange Eq.~\eqref{eq:WallMomentumFlux} in terms of the $x$ component slip velocity:
	\begin{multline}
		u_{s}=\pm s\left(1+\cos{\theta_{0}}\right)\times\\
		\frac{\left(1-\frac{1}{2}\sin^{2}{\theta_{0}}\frac{\rho_{0}d_{0}T^{1/2}g_{w}}{\eta}\left\{\frac{5\sqrt{\pi}}{96g_{0}}\left(1+\frac{8}{5}\phi g_{0}\right)+\frac{\sqrt{2}\phi}{\sqrt{\pi}}\right\}\right)}{\frac{2\sqrt{2}}{3\sqrt{\pi}}\rho_{0}\phi g_{w}T^{1/2}\left(2-\left(1+\cos{\theta_{0}}\right)\cos{\theta_{0}}\right)},
		\label{eq:NottSlip}
	\end{multline}
where the positive and negative signs correspond to the lower and upper walls, respectively.
We determine $\theta_{0}$ by fitting Eq.~\eqref{eq:NottSlip} to our \ac{DEM} results, using measured shear stress and slip velocity at the wall, as shown in Fig.~\ref{fig:theta0fit}.
For simplicity, we fit a single mean value of $\theta_{0}=(0.67\pm0.02)\SI{}{rad}$ for all the cases, which is approximately equal to the value of $\pi/5$ obtained analytically in other works \cite{VescoviGranularShearCouetteDEM2014}.
\par
Temperature is usually constrained in rarefied gas modelling through a first-order ``jump'' boundary condition \cite{LockerbyBurnettMicroCouette2003,DadzieBivelocityGasLidCavity2016,ChristouHeatTransferCavity2017}, or alternatively, the case may be assumed isothermal to simplify the analyses \cite{ChristouRarefiedGasRecasted2018,TomyDiffusionSlip2022}. However, in a DEM simulation, temperature is not trivial to determine for a rough granular wall, and the ``thermostatting'' approaches often used in \ac{MD} simulations do not work well in \ac{DEM}. Here we apply the wall heat flux model boundary conditions from \citet{NottBCWallGranular2011}:
\begin{equation}
	\hat{\boldsymbol{n}}\cdot\boldsymbol{q}=\Upsilon_{w}-\boldsymbol{\Lambda}_{w}\cdot\boldsymbol{u}_{s},
	\label{eq:NottHeatFlux}
\end{equation}
where:
\begin{equation}
	\Upsilon_{w}=\left(\frac{2}{\pi}\right)^{1/2}\left(1+\cos{\theta_{0}}\right)^{-1}\left(1-e^{2}\right)\rho_{0}\phi g_{w}T^{3/2},
	\label{eq:Dissipation}
\end{equation}
is the rate of kinetic energy dissipation at the wall per unit area due to inelastic collisions.
For our case, we simplify Eq.~\eqref{eq:NottHeatFlux}, and equate Eq.~\eqref{eq:WallMomentumFlux} to the normal stress at the wall, $S_{yy}$, to obtain two sets of boundary conditions, for heat flux, $q_{w}$, and volume diffusion flux, $J_{c,w}$, respectively, at the walls:
\begin{equation}
	q_{w}=\pm\left(-\Upsilon_{w}+s\left|u_{s}\right|\right),
	\label{eq:heatfluxBC0}
\end{equation}
and
\begin{equation}
	J_{c,w}=\frac{\pm\left(S_{yy}-\rho_{0}\phi{}g_{w}T\right)+\frac{A}{\kappa}q_{w}}{\frac{A\mu\phi}{\kappa k_{m}}-\frac{B\phi}{k_{m}}},
	\label{eq:JcwfluxBC}
\end{equation}
where the positive and negative signs of $(\pm)$ correspond to the lower and upper boundaries, respectively, and coefficients $A$ and $B$ are derived from the tensors in Eq.~\eqref{eq:NottTensors}:
\begin{equation}
	A=\rho_{0}d_{0}g_{w}\left[\frac{5}{96\sqrt{2}g_{0}}\left(1+\frac{12}{5}G\right)+\frac{\phi}{3}\right]\frac{2\left(1-\cos^{3}{\theta_{0}}\right)}{\sin^{2}{\theta_{0}}},
	\label{eq:WallCoeffA}
\end{equation}
and
\begin{equation}
	B=\frac{1}{3}\rho_{0}d_{0}Tg_{w}\frac{2\left(1-\cos^{3}{\theta_{0}}\right)}{\sin^{2}{\theta_{0}}},
	\label{eq:WallCoeffB}
\end{equation}
respectively.
\par
Additionally, Eq.~\eqref{eq:HoldupDiff} has boundary conditions \cite{VescoviGranularShearCouetteDEM2014}:
\begin{subequations}
	\begin{equation}
		m(y=0)=0,
	\end{equation}
	\begin{equation}
		m(y=H)=\bar{\phi}H,
	\end{equation}
	\label{eq:MassHoldUpBC}
\end{subequations}
which satisfies the mean volume fraction in the channel. Eqs.~\eqref{eq:NottSlip}, \eqref{eq:JcwfluxBC}, \eqref{eq:heatfluxBC0}, and \eqref{eq:MassHoldUpBC} now provide sufficient boundary conditions to fully solve for our \ac{VD} model in Eqs.~\eqref{eq:VelocityDiff}--\eqref{eq:HoldupDiff}, with additional parameters $s$ and $S_{yy}$.

\section{Results and Discussions} \label{sec:ResultsDiscussions}
We solved Eqs.~\eqref{eq:VelocityDiff}--\eqref{eq:HeatFluxDiff} and \eqref{eq:HoldupDiff}, with boundary conditions defined by Eqs.~\eqref{eq:NottSlip}, \eqref{eq:heatfluxBC0}--\eqref{eq:MassHoldUpBC}, using the \textsc{BVP4C} solver function in \textsc{MATLAB} \cite{MATLAB:R2021a_u1}, and compared results to our \ac{DEM} simulations. For validations, we determined the conventional \ac{NSF} solution, by solving Eqs.~\eqref{eq:VelocityDiff}, and \eqref{eq:VoidDiffNS}--\eqref{eq:HoldupDiff}, while using the same rough wall boundary conditions in Eqs.~\eqref{eq:NottSlip}, \eqref{eq:heatfluxBC0}, and \eqref{eq:MassHoldUpBC}.
\par
\begin{figure*}[!htb]
	\includegraphics{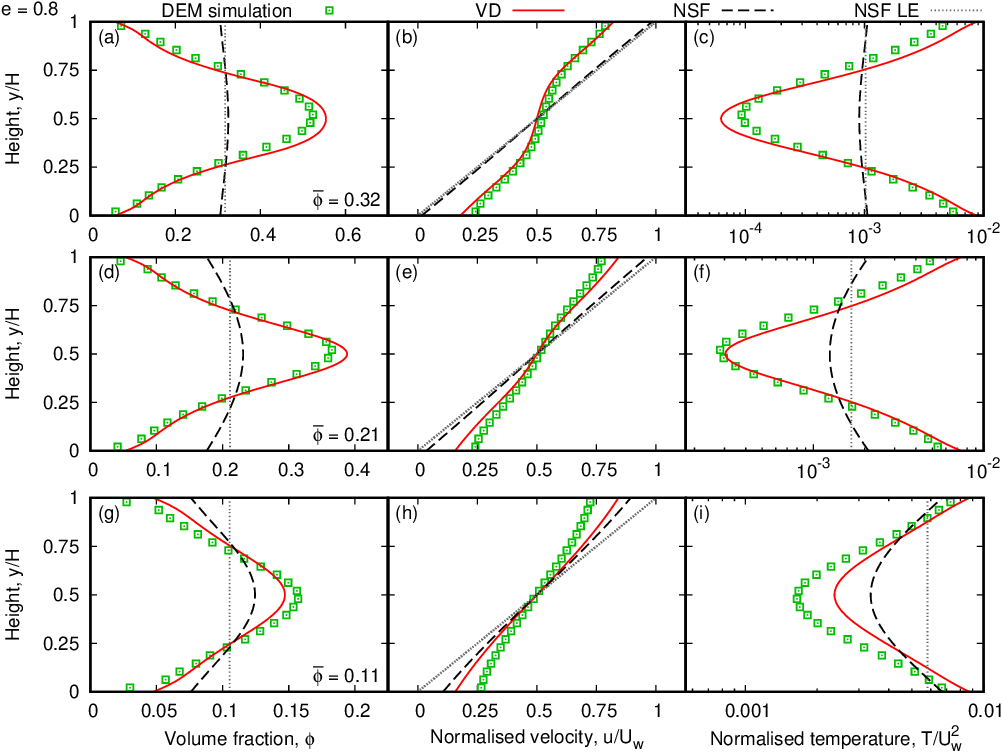}
	\caption{Variation in volume fraction, $\phi$, velocity, $u/U_{w}$, and temperature, $T/U_{w}^{2}$, in the: (a)--(c)~$\bar{\phi}=\SI{0.32}{}$, (d)--(f)~$\bar{\phi}=0.211$, and (g)--(i)~$\bar{\phi}=\SI{0.11}{}$ cases, all for $e=0.8$.}
	\label{fig:ChannelProfiles}
\end{figure*}

Fig.~\ref{fig:ChannelProfiles} shows the variations in volume fraction, $\phi$, normalised velocity, $u/U_{w}$, and normalised temperature, $T/U_{w}^{2}$, across the channel for the $\bar{\phi}=0.32,\ \SI{0.211}{}$, and $\SI{0.11}{}$ cases, where the \ac{VD} model gives a general improvement over the \ac{NSF} model.\footnote{For the remainder of this article, results for the velocity, temperature, volume diffusion flux, and stresses are normalised by $U_{w}$, $U_{w}^{2}$, $U_{w}$, and $\rho_{0}U_{w}^{2}(d_{0}/H)^{2}$, respectively.}
The \ac{VD} equations typically result in higher peak volume fractions near the centreline ($y/H=0.5$), greater slip velocities, and lower granular temperatures.
In Fig.~\ref{fig:ChannelProfiles}, we have also plotted the results predicted by the \acl{NSF} models under \acf{LE} boundary conditions \cite{LeesEdwardsBC1972}, where shear flow is imposed through the displacement of the $y$ periodic faces \cite{KlossLIGGGHTS2012}, and no rough walls. These are solved using the same differential equations as the \ac{NSF} solution, however, assuming no slip and no heat transfer at the boundaries (adiabatic walls), i.e.\ $u_{s}=0$ and $q_{w}=0$, respectively, which are also the assumptions in models proposed by \citet{LunKineticGranularCouette1984} and \citet{ChialvoKineticFrictionGranular2013}.
\par
\begin{figure*}[!htb]
	\includegraphics{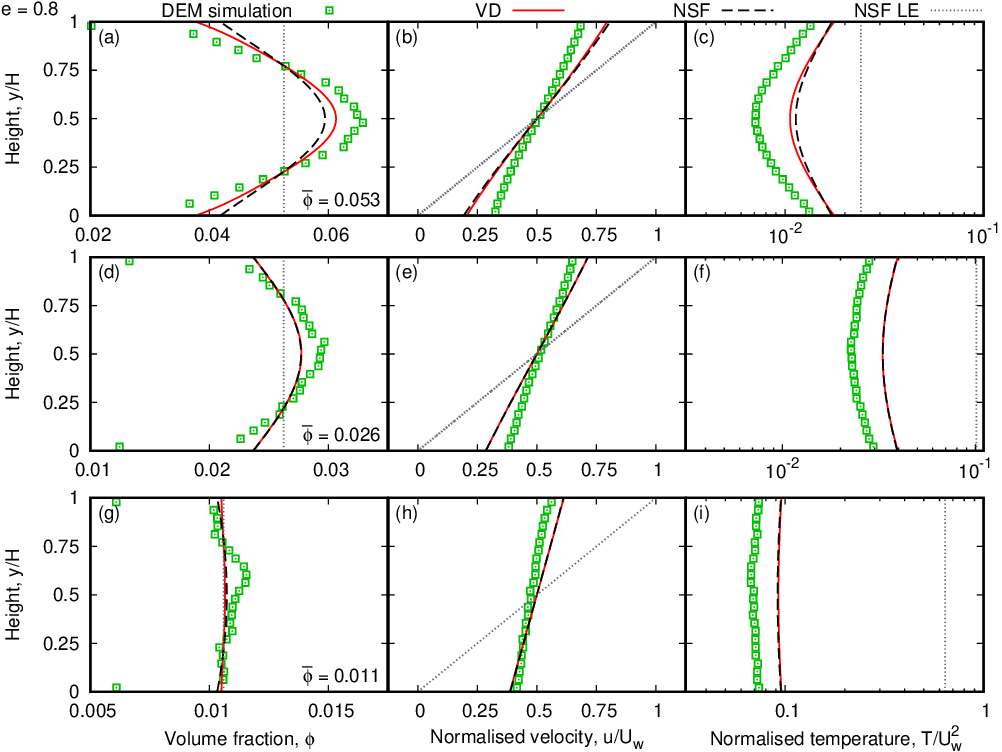}
	\caption{Variation in volume fraction, $\phi$, velocity, $u/U_{w}$, and temperature, $T/U_{w}^{2}$, in the: (a)--(c)~$\bar{\phi}=\SI{0.053}{}$, (d)--(f)~$\bar{\phi}=0.026$, and (g)--(i)~$\bar{\phi}=\SI{0.011}{}$ cases, all for $e=0.8$.}
	\label{fig:ChannelProfilesLowDensity}
\end{figure*}

Our \ac{VD} model generally provides better agreement over the \ac{NSF} models for the above cases with $\bar{\phi}\sim0.1$, which occurs since there are large volume fraction gradients, which means the volume diffusion flux term, $J_{c}$, becomes more significant, from Eq.~\eqref{eq:Jc}. For lower volume fraction cases $\bar{\phi}\lessapprox0.05$, the volume fraction profiles became more uniform across the channel wall, and the \ac{VD} and \ac{NSF} solutions tend to show similar results, as shown in Fig.~\ref{fig:ChannelProfilesLowDensity}.
\citet{AlamInstabilityOrderingGravity2005} found qualitatively similar results, albeit assuming adiabatic walls: for low $\bar{\phi}$, the volume fraction across the channel is uniform, however, as $\bar{\phi}$ increases, layering instabilities dominate the flow profile, giving rise to the peak density variations \cite{AlamStabilityCouetteFlow1998,NottBoundariesCouetteBifurcation1999,AlamInstabilityOrderingGravity2005}, which we see in our \ac{DEM} simulations, and this is consistent with our \ac{VD} model which shows the greatest deviations with the \ac{NSF} model for higher mean volume fractions $\bar{\phi}\sim0.1$.
\par
\begin{figure*}[!htb]
	\begin{center}
		\includegraphics{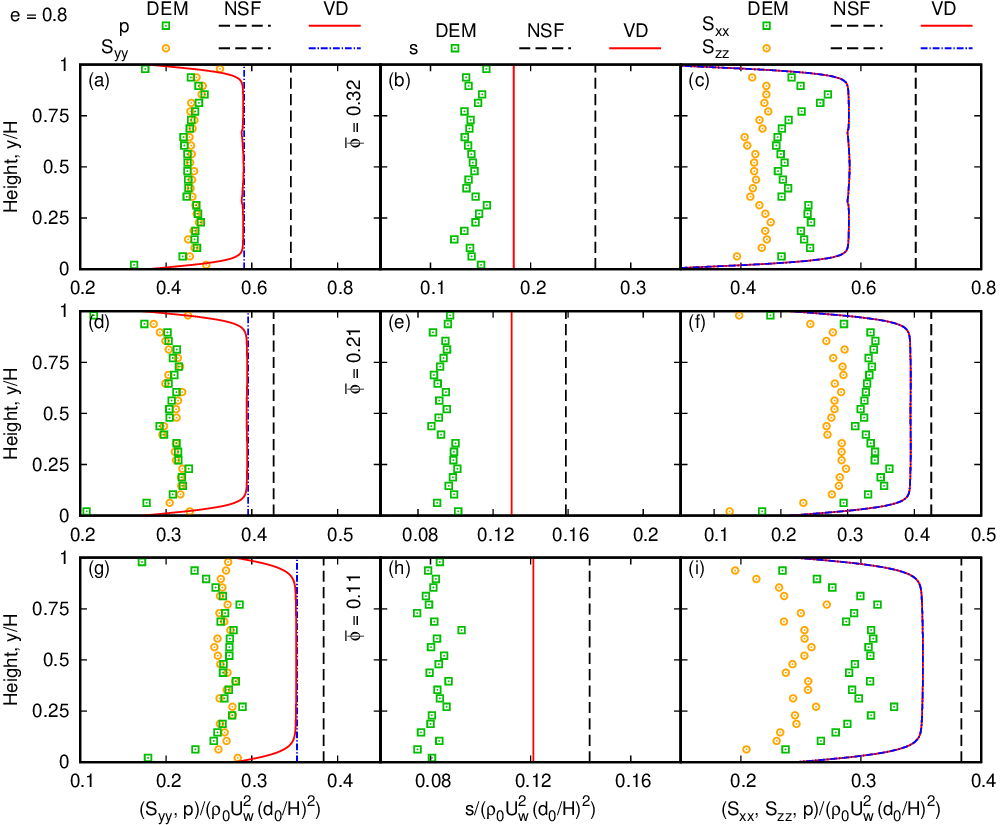}
		\caption{Variation in normal stress component, $S_{yy}$ and pressure, $p$, shear stress, $s$, and $x$ and $z$ normal stresses, $S_{xx}$ and $S_{zz}$, respectively, in the: (a)--(c)~$\bar{\phi}=\SI{0.32}{}$, (d)--(f)~$\bar{\phi}=0.21$, and (g)--(i)~$\bar{\phi}=\SI{0.11}{}$ cases, all for $e=0.8$.
		The legends above (a), (b), and (c) apply to all subfigures in the same column.}
		\label{fig:ChannelProfilesStress}
	\end{center}
\end{figure*}

Our \ac{VD} model also better predicts the stresses for these cases, compared to the classical \acl{NSF} equations, as shown in Fig.~\ref{fig:ChannelProfilesStress}, for various volume fractions at $e=0.8$.
The normal stress, $S_{yy}$, and pressure, $p$, are measured in the \ac{DEM} results using Eqs.~\eqref{eq:VirialStressTensor} and \eqref{eq:PDEM}, respectively. Crucially, we can see the non-uniform variation in pressures in the \ac{DEM} and \ac{VD} model results in Figs.~\ref{fig:ChannelProfilesStress}(a), (d), and (g), which are not captured by the conventional \ac{NSF} equations. Instead, it is the normal $y$-stress tensor component, $S_{yy}$, and the shear stress $s$ in Figs.~\ref{fig:ChannelProfilesStress}(b), (e), and (h), which remain constant across the channel as expected from Eqs.~\eqref{eq:pressuregradient0} and \eqref{eq:VelocityDiff}, respectively. To further demonstrate the difference between measured stress tensor and pressure profiles, we have plotted the normal $x$ and $z$ stress tensor components in Figs.~\ref{fig:ChannelProfilesStress}(c), (f), and (i), obtained from the thermodynamic pressure and the stress tensor elements in Eq.~\eqref{eq:VDstresstensorreduced}:
\begin{equation}
	S_{xx}=S_{zz}=p+\left(-\frac{2\eta}{3}+\eta_{b}\right)\frac{\partial J_{c}}{\partial y}.
\end{equation}
Again, we show clearly non-uniform variations in the stresses $S_{xx}$ and $S_{zz}$, while the \ac{NSF} solution predicts constant values (since there is no volume diffusion flux terms).
However, contrary to what we expected from the \ac{VD} stress tensor in Eq.~\eqref{eq:VDstresstensorreduced}, the $S_{xx}$ and $S_{zz}$ stresses are not exactly equal in our \ac{DEM} simulations, with the normal $x$ component consistently larger than the $z$ component by around $0.1$ (normalised by $\rho_{0}U_{w}^{2}(d_{0}/H)^{2}$). This is likely because we only considered volume fraction gradients in Eq.~\eqref{eq:Jc}, however, other gradient terms, such as in thermodynamic pressure and temperature, are also theorised to contribute to the volume diffusion flux \cite{LakshminarayanaRNS2019,StamatiouEnhancedFlowNanotube2019,TomyDiffusionSlip2022}, and these quantities are clearly non-uniform in our cases here. Introducing more volume diffusion flux terms is challenging, as we would also have to introduce more necessary boundary conditions to close the problem, so for the purposes of this work, we limited ourselves to capturing volume diffusion flux driven by volume fraction gradients only.
\par
\par
\begin{figure}[!htb]
	\begin{center}
		\includegraphics[width=0.5\textwidth]{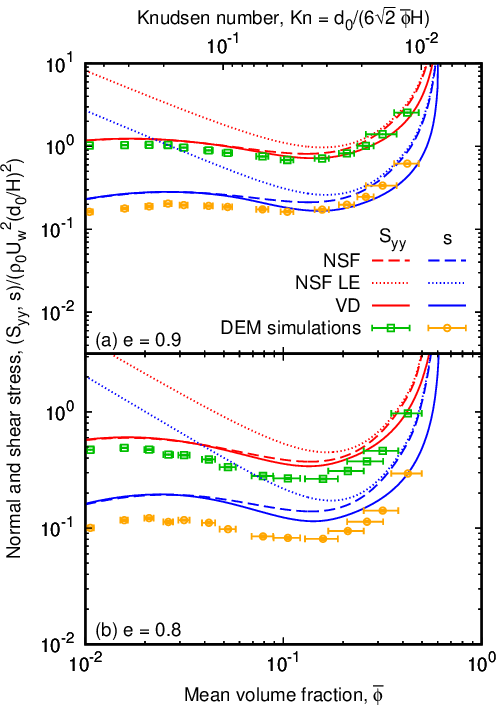}
		\caption{Variation in normal stress, $S_{yy}$ and shear stress, $s$, for the \ac{NSF} and \ac{VD} and \ac{NSF} (with \ac{LE} boundary condition) models, for the (a)~$e=0.9$ and (b)~$e=0.8$ cases. The Knudsen number, $\Knn$, applies only to the \ac{NSF} and \ac{VD} channel models, i.e.\ with rough walls. The legend in (a) also applies to (b) and (c).}
		\label{fig:StressVoid}
	\end{center}
\end{figure}

In Fig.~\ref{fig:StressVoid}(a)--(c), we show the variation in stresses against volume fraction for the \ac{VD}, \ac{NSF}, and \ac{NSF} with \ac{LE} boundary condition models, for $e=0.9$ and $0.8$, respectively.
The \ac{NSF} model captures $s$ and $S_{yy}$ at low ($\bar{\phi}\sim0.01$) volume fractions, and is in generally good agreement with our \ac{VD} model. However, as we discussed earlier, for $\bar{\phi}\gtrapprox0.05$, the conventional \ac{NSF} solutions deviate from our \ac{VD} model, which occurs for cases where there were large density gradients (see Fig.~\ref{fig:ChannelProfiles}), most prominently at $\bar{\phi}\sim0.1$, and volume diffusion flux becomes relevant. This disagreement continues for further increases of $\bar{\phi}$, until we reach the maximum random close packing limit $\phi_{rcp}$, which is the maximum physical case we can explore in either the \ac{NSF} or \ac{VD} models.
\par
As we discussed above, the \ac{VD} and \ac{NSF} equations converged on similar solutions for $\bar{\phi}\sim0.01$, since the volume fraction profiles became more uniform across the channel. This explains a unique finding in our \ac{VD} modelling, in comparison with previous applications of \ac{VD} theory. \ac{VD} equations, and general bi-velocity methods, have previously been used to model rarefied gas dynamics, generally showing greater effects as $\Knn$ increases, typically above unity, for low-density cavity flows and pressure-driven micro-channel flows \cite{DadzieBivelocityGasLidCavity2016,ChristouHeatTransferCavity2017,ChristouRarefiedGasRecasted2018}.
In our current simulations, we express Knudsen numbers in the configurations as $\Knn=d_{0}/(6\sqrt{2}\bar{\phi}H)$ \cite{LockerbyBurnettMicroCouette2003,DadzieBivelocityGasLidCavity2016,ChristouHeatTransferCavity2017,ChristouRarefiedGasRecasted2018} which applies to the \ac{NSF} and \ac{VD} channel models only, where the flow is confined between two walls. However, the \ac{NSF} and \ac{VD} equations show almost complete agreement as $\Knn$ approaches unity, and instead the largest deviation occurs at $\Knn\sim0.01$, or approximately $\bar{\phi}\sim0.1$, where layering instabilities result in a peak density variation, which develop the density gradient dependent volume diffusion flux terms in the \ac{VD} equations. We conclude that in this granular shearing case, the Knudsen number does \textit{not} play a significant role after all, and we should focus on cases where volume fraction gradients are more significant, which might not necessarily occur in rarefied flows.
\par
Despite the deviations between the \ac{VD} and \ac{NSF} models, Fig.~\ref{fig:StressVoid} shows that both models give similar qualitative agreements with the \ac{DEM} results, when looking at normal and shear stresses, however, as we showed previously in Figs.~\ref{fig:ChannelProfiles} and \ref{fig:ChannelProfilesStress}, the \ac{VD} model is able to provide better detail for the non-uniform varying properties within the channel, such as volume fraction and temperature, and also thermodynamic pressure variations which the \ac{NSF} model cannot capture.
\par
The differences between the \ac{NSF} and \ac{VD} models seem weakly dependent on the coefficient of restitution, as shown in Figs.~\ref{fig:StressVoid}(a)--(c) for $e=0.9$ and $0.8$ cases, respectively. For each value of $e$, we see close agreement between the \ac{NSF} and \ac{VD} models at low volume fraction $\bar{\phi}\sim0.01$, with deviations between the two models generally increasing as $\bar{\phi}$ increases, due to the larger volume fraction gradients. This suggests that any discrepancies in the \ac{NSF} for shear flow modelling are less dependent on particle inelasticity, so long as a peak volume fraction variation occurs, which still occurs in elastic particle interactions \cite{LockerbyBurnettMicroCouette2003}.
\par
We find that both the \ac{NSF} and \ac{VD} tend to overpredict stresses in the \ac{DEM} simulations for the low volume fraction cases $\bar{\phi}\sim0.01$, which becomes more apparent as $e$ decreases, however, we also find that the \ac{NSF} model with \ac{LE} boundary conditions also overpredicts stresses in this range, as shown in Appendix~\ref{Appendix:LeesEdwardsBC}. From this additional comparison, we suggest that the disagreement most likely arises from the fluid transport coefficients used here, rather than our use of \ac{VD} modelling or choice of boundary conditions in \S\ref{sec:VDmodel}.
\par
We also note that the value of $\theta_{0}$ in Fig.~\ref{fig:theta0fit}, used to determine slip velocity in Eq.~\eqref{eq:NottSlip}, and energy dissipation rate at the wall in Eq.~\eqref{eq:Dissipation}, appeared to decrease with increasing volume fraction, which could have affected agreement with our \ac{DEM} results. It may seem unusual that $\theta_{0}$ is not necessarily constant, as one might assume $\theta_{0}$ is a purely geometric parameter, relating to the wall's \ac{FCC} structure \cite{RichmanBC1988,VescoviGranularShearCouetteDEM2014,NottBCWallGranular2011}. However, the coefficient of restitution and volume fraction have previously been shown to alter the mean collision-angle distribution in inelastic particle collisions \cite{CampbellBrennenSimulationGranular1985}, and we suggest this could also be occurring at the walls in our cases. We expect for the dilute ($\bar{\phi}\rightarrow0$) cases, $\theta_{0}$ would tend to a constant value; however, for increasing local volume fraction, dense layered microstructures form \cite{CampbellBrennenSimulationGranular1985}, which break down the assumption of molecular chaos, and the collision angle at the wall $\theta_{0}$ would naturally become dependent on the surrounding high density of particles.

\section{Conclusions} \label{sec:Conclusions}
\acresetall
We have adapted the \ac{VD} extended hydrodynamic equations to model the shearing of a granular flow confined between two walls, using gas properties obtained from kinetic theory for inelastic particles \cite{GarzoDuftyGranularKinetic1999}. We compared our new \ac{VD} model to \ac{DEM} simulations for granular flow contained between two rough shearing walls, where we varied mean volume fraction $\bar{\phi}$ and the inelasticity of inter-particle collisions, through the coefficient of restitution $e$. The \ac{VD} equations included a volume fraction (or normalised density) gradient term, the \textit{volume diffusion flux}, $J_{c}$, which better captured the volume fraction and granular temperature profiles across the channel, when compared to the classical \ac{NSF} equations. We employed boundary conditions for slip-velocity and heat flux, commonly used in granular flow literature \cite{NottBCWallGranular2011}, and derived a new set of boundary conditions for the volume diffusion flux, by equating the normal stress at the upper and lower walls with the \ac{VD} stress tensor.
Our \ac{VD} model provided good agreement with \ac{DEM} simulations for predicting local volume fraction, flow velocity, and granular temperature, and was able to capture non-uniform variations in pressure and stresses, unlike the classical \ac{NSF} equations.
\par
Contrary to what we expected from previous investigations \cite{DadzieBivelocityGasLidCavity2016,ChristouHeatTransferCavity2017,ChristouRarefiedGasRecasted2018}, we found that \ac{VD} equations were weakly dependent on Knudsen number $\Knn$ approaching unity and beyond, giving very similar results to \ac{NSF} equations. This results from low the volume fraction cases producing near-uniform density across the channel for $\bar{\phi}\lessapprox0.05$ \cite{AlamStabilityCouetteFlow1998,NottBoundariesCouetteBifurcation1999,AlamInstabilityOrderingGravity2005}, as found in our \ac{DEM} simulations. Increasing the mean volume fraction in our simulations introduced layering instabilities for $\bar{\phi}\sim0.1$, which resulted in peak volume fraction variations in the channel, and where the volume diffusion flux terms became most significant, resulting in the regime where we saw the greatest deviations between our \ac{VD} model and the classical \ac{NSF} equations.
\par
Our derived \ac{VD} model only considered the volume diffusion flux from density gradients, however, gradients in temperature and thermodynamic pressure can also induce volume diffusion flux terms, that could account for some of discrepancies in our normal and shear stress results \cite{LakshminarayanaRNS2019,StamatiouEnhancedFlowNanotube2019,TomyDiffusionSlip2022}. Adapting the model to account for these additional \ac{VD} flux terms would also require more necessary boundary conditions, and likely modifications to the existing velocity slip and heat flux boundary conditions, which may partly explain the variation of particle-wall collision angle $\theta_{0}$ with local volume fraction we observed in our \ac{DEM} simulation results.
We assumed perfectly smooth particles whereas more physically relevant granular systems in nature and engineering often contain rough particles, with finite inter-particle friction, and non-uniform size. These can be implemented for better applications for the experimental validations \cite{NottBCWallGranular2011,ChialvoKineticFrictionGranular2013,AbuZaidKineticGranularFrictional1990,LunKineticGranularRoughInelastic1987}.
Future work could also further investigate the application of the \ac{VD} equations for modelling asymmetric flow behaviour, which becomes more prevalent at lower values of $e$ \cite{AlamStabilityCouetteFlow1998,NottBoundariesCouetteBifurcation1999,AlamInstabilityOrderingGravity2005}.
\section*{Acknowledgements}
This work is supported in the UK by the Engineering and Physical Sciences Research Council (EPSRC) under grant EP/R007438/1.

\begin{appendix}
\setcounter{figure}{0}
\setcounter{equation}{0}
\setcounter{table}{0}
\renewcommand\thefigure{\thesection\arabic{figure}}
\renewcommand\thetable{\thesection\arabic{table}}
\renewcommand\theequation{\thesection\arabic{equation}}
\section{Transport coefficients for a granular gas} \label{Appendix:Coefficients}
Table~\ref{tab:Coefficients} gives the transport coefficients, approximating a granular medium as a gas \cite{GarzoDuftyGranularKinetic1999}, and using the radial distribution function defined in Eqs.~\eqref{eq:RDF} and \eqref{eq:VescoviFactor}.
Coefficients denoted by the asterisk ($^{*}$) superscript are given in dimensionless form, and can be related by:
\begin{subequations}
	\begin{gather}
	\eta^{*}=\eta/\eta_{0},\\
	\eta_{b}^{*}=\eta_{b}/\eta_{0},\\
	\kappa^{*}=\kappa/\kappa_{0},\\
	\mu^{*}=\mu\phi/T\kappa_{0},\\
	\zeta^{*}=\Gamma/(3\rho_{0}\phi T\nu_{0}/2),
	\end{gather}
\end{subequations}
where:
\begin{equation}
	\eta_{0}=\frac{5}{96}\rho_{0}\pi^{1/2}d_{0}T^{1/2},
\label{eq:viscosity0}
\end{equation}
and:
\begin{equation}
	\kappa_{0}=\frac{15}{4}\eta_{0},
	\label{eq:kappa0}
\end{equation}
are the low-volume fraction values of shear viscosity and thermal conductivity, respectively, in the elastic limit ($e=1$) \cite{GarzoDuftyGranularKinetic1999}. Additionally, the characteristic collision frequency is given by:
\begin{equation}
	\nu_{0}=\rho_{0}\phi T/\eta_{0}.
\end{equation}

\begin{table*}[!htb]
	\begin{threeparttable}
		\caption{Fluid transport coefficients for a granular gas \cite{GarzoDuftyGranularKinetic1999,NottBCWallGranular2011}.}
		\label{tab:Coefficients}
		\begin{tabular}{p{\textwidth}}
			\hline
			\begin{equation*}
				G=\phi g_{0},
			\end{equation*}
			\begin{equation*}
				g_{w}=1+4G, 
			\end{equation*}
			\begin{equation*}
				F=\frac{1+e}{2}+\frac{1}{4G},
			\end{equation*}
			\begin{equation*}
				\eta^*= \eta^{k *}\left[1+\frac{4}{5} G\left(1+e\right)\right]+\frac{3}{5} \eta_{b}^*, 
			\end{equation*}
			\begin{equation*}
				\eta^{k *}=\left(\nu_\eta^*-\frac{1}{2} \zeta^{*}\right)^{-1}\left[1-\frac{2}{5}G(1+e)(1-3 e) \right],
			\end{equation*}
			\begin{equation*}
				c^*= 32(1-e)\left(1-2 e^2\right)\left[81-17 e+30 e^2(1-e)\right]^{-1}
			\end{equation*}
			\begin{equation*}
				\eta_{b}^*= \frac{128}{5\pi} \phi G(1+e)\left(1-\frac{1}{32} c^*\right),
			\end{equation*}
			\begin{equation*}
				\kappa^*= \kappa^{k *}\left[1+\frac{6}{5} G(1+e)\right]+\frac{2304}{225\pi} \phi G(1+e)\left(1+\frac{7}{32} c^*\right),
			\end{equation*}
			\begin{equation*}
				\kappa^{k *}= \frac{2}{3}\left(\nu_{\kappa}^{*}-2 \zeta^{*}\right)^{-1}\left(1+\frac{1}{2}\left(1+4FG \right) c^{*}
				+\frac{3}{5} G(1+e)^2\left\{2 e-1+\left[\frac{1}{2}(1+e)-\frac{5}{3(1+e)}\right] c^*\right\}\right)
			\end{equation*}
			\begin{equation*}
				\mu^*= \mu^{k *}\left[1+\frac{6}{5} G(1+e)\right],
			\end{equation*}
			\begin{multline*}
				\mu^{k *}= 2\left(2 \nu_{\mu}^{*}-3 \zeta^{*}\right)^{-1}
				\left[\left(1+\phi \partial_\phi \ln g_{0} \right) \zeta^{*} \kappa^{k *}+\frac{4FG }{3}\left(1+\phi \partial_\phi \ln {\left\{4FG\right\}} \right) c^*\right.\\
\left.-\frac{4}{5} G\left(1+\frac{1}{2} \phi \partial_\phi \ln g_{0} \right)(1+e)\left\{e(1-e)+\frac{1}{4}\left[\frac{4}{3}+e(1-e)\right] c^*\right\}\right],
			\end{multline*}
			\begin{equation*}
				\zeta^{*}=\frac{5}{12} g_{0} \left(1-e^2\right)\left(1+\frac{3}{32} c^*\right), \tnote{a}
			\end{equation*}
			\begin{equation*}
				\nu_\eta^*= g_{0} \left[1-\frac{1}{4}(1-e)^2\right]\left[1-\frac{1}{64} c^*\right],
			\end{equation*}
			\begin{equation*}
				\nu_\kappa^*= \nu_\mu^*=\frac{1}{3}(1+e) g_{0} \left[1+\frac{33}{16}(1-e)+\frac{19-3 e}{1024} c^*\right]
			\end{equation*}
			\\
			\hline
		\end{tabular}
		\begin{tablenotes}
			\item[a] We have neglected the divergence term in the dimensionless dissipation rate $\zeta^{*}$, from Ref.~\citenum{GarzoDuftyGranularKinetic1999}.\hfill\break
		\end{tablenotes}
	\end{threeparttable}
\end{table*}

\section{Lees--Edwards boundary conditions} \label{Appendix:LeesEdwardsBC}
\setcounter{figure}{0}
In this appendix, we show results of our \ac{DEM} simulations under \ac{LE} boundary conditions, which were set up in a fully periodic box, with dimensions $\SI{0.5}{m}$, $\SI{0.25}{m}$, and $\SI{0.05}{m}$, in the $x$, $y$, and $z$ dimensions, respectively, and run for $\SI{e7}{}\Delta t$ to reach a steady-state and $\SI{e8}{}\Delta t$ for averaging.
Eq.~\eqref{eq:RDF} provides a better fit than the \ac{CS} radial distribution function \cite{CarnahnStarlingg01969}, particularly at $\bar{\phi}>0.4$, as shown in Fig.~\ref{fig:RDFfit}.
\begin{figure}[!htb]
	\begin{center}
		\includegraphics[width=0.5\textwidth]{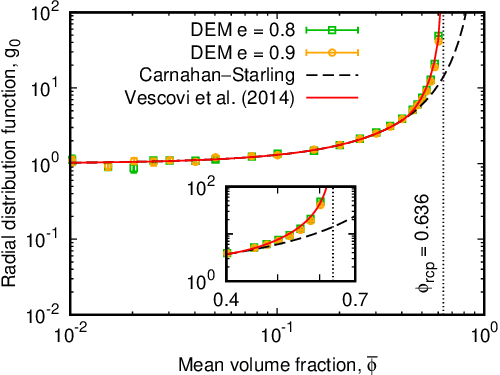}
		\caption{Variation in radial distribution function, $g_{0}$, with mean volume fraction, $\bar{\phi}$, from the \ac{DEM} simulations, and predicted using Eqs.~\eqref{eq:RDF} and \eqref{eq:VescoviFactor} proposed by \citet{VescoviGranularShearCouetteDEM2014}, with comparisons to the Carnahan--Starling radial distribution function \cite{CarnahnStarlingg01969}. The dotted line shows the maximum possible volume fraction, at the random close packing limit $\phi_{rcp}$, and the inset shows results for higher volume fractions $\phi>0.4$ in more detail.}
		\label{fig:RDFfit}
	\end{center}
\end{figure}
We also see no significant effect of $e$ on $g_{0}$.
\par
\begin{figure}[!htb]
	\begin{center}
		\includegraphics[width=0.5\textwidth]{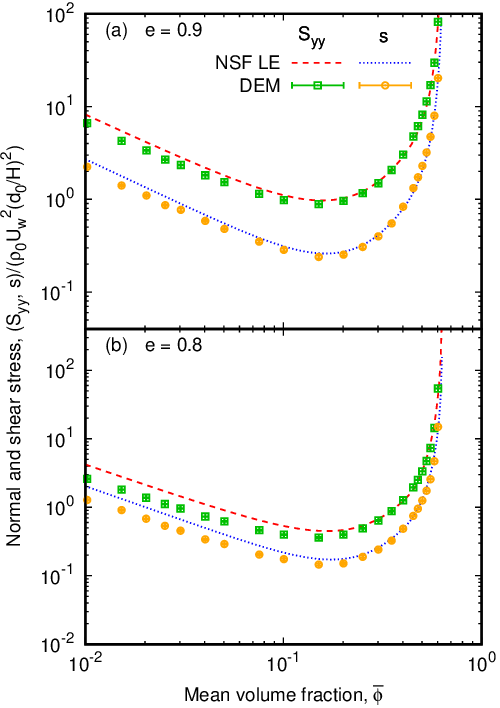}
		\caption{Variation in the normal and shear stresses, $S_{yy}$ and $s$, respectively, for simulations run under \ac{LE} boundary conditions, for coefficients of restitution: (a)~$e=0.9$ and (b)~$e=0.8$.}
		\label{fig:StressVoidLE}
	\end{center}
\end{figure}

Fig.~\ref{fig:StressVoidLE} shows the variation in normal and shear stresses inside the channel under \ac{LE} boundary conditions, which are in good agreement with the \ac{NSF} equations.
Since no density fluctuations were present, due to the absence of walls in \ac{LE} boundary conditions, there is no difference between the \ac{NSF} and \ac{VD} equations solutions, and therefore $p=S_{yy}$.
\end{appendix}
\FloatBarrier

\bibliographystyle{apsrev4-1}

\begin{thebibliography}{51}%
\makeatletter
\providecommand \@ifxundefined [1]{%
 \@ifx{#1\undefined}
}%
\providecommand \@ifnum [1]{%
 \ifnum #1\expandafter \@firstoftwo
 \else \expandafter \@secondoftwo
 \fi
}%
\providecommand \@ifx [1]{%
 \ifx #1\expandafter \@firstoftwo
 \else \expandafter \@secondoftwo
 \fi
}%
\providecommand \natexlab [1]{#1}%
\providecommand \enquote  [1]{``#1''}%
\providecommand \bibnamefont  [1]{#1}%
\providecommand \bibfnamefont [1]{#1}%
\providecommand \citenamefont [1]{#1}%
\providecommand \href@noop [0]{\@secondoftwo}%
\providecommand \href [0]{\begingroup \@sanitize@url \@href}%
\providecommand \@href[1]{\@@startlink{#1}\@@href}%
\providecommand \@@href[1]{\endgroup#1\@@endlink}%
\providecommand \@sanitize@url [0]{\catcode `\\12\catcode `\$12\catcode
  `\&12\catcode `\#12\catcode `\^12\catcode `\_12\catcode `\%12\relax}%
\providecommand \@@startlink[1]{}%
\providecommand \@@endlink[0]{}%
\providecommand \url  [0]{\begingroup\@sanitize@url \@url }%
\providecommand \@url [1]{\endgroup\@href {#1}{\urlprefix }}%
\providecommand \urlprefix  [0]{URL }%
\providecommand \Eprint [0]{\href }%
\providecommand \doibase [0]{http://dx.doi.org/}%
\providecommand \selectlanguage [0]{\@gobble}%
\providecommand \bibinfo  [0]{\@secondoftwo}%
\providecommand \bibfield  [0]{\@secondoftwo}%
\providecommand \translation [1]{[#1]}%
\providecommand \BibitemOpen [0]{}%
\providecommand \bibitemStop [0]{}%
\providecommand \bibitemNoStop [0]{.\EOS\space}%
\providecommand \EOS [0]{\spacefactor3000\relax}%
\providecommand \BibitemShut  [1]{\csname bibitem#1\endcsname}%
\let\auto@bib@innerbib\@empty
\bibitem [{\citenamefont {Friedmann}\ \emph {et~al.}(2006)\citenamefont
  {Friedmann}, \citenamefont {Taberlet},\ and\ \citenamefont
  {Losert}}]{FriedmannRockAvalanches2006}%
  \BibitemOpen
  \bibfield  {author} {\bibinfo {author} {\bibfnamefont {S.~J.}\ \bibnamefont
  {Friedmann}}, \bibinfo {author} {\bibfnamefont {N.}~\bibnamefont {Taberlet}},
  \ and\ \bibinfo {author} {\bibfnamefont {W.}~\bibnamefont {Losert}},\ }\href
  {\doibase 10.1007/s00531-006-0067-9} {\bibfield  {journal} {\bibinfo
  {journal} {Int. J. Earth Sci.}\ }\textbf {\bibinfo {volume} {95}},\ \bibinfo
  {pages} {911} (\bibinfo {year} {2006})}\BibitemShut {NoStop}%
\bibitem [{\citenamefont {Cuomo}(2020)}]{CuomoModellingFlowslidesReview2020}%
  \BibitemOpen
  \bibfield  {author} {\bibinfo {author} {\bibfnamefont {S.}~\bibnamefont
  {Cuomo}},\ }\href {\doibase 10.1186/s40677-019-0133-9} {\bibfield  {journal}
  {\bibinfo  {journal} {Geoenvironmental Disasters}\ }\textbf {\bibinfo
  {volume} {7}},\ \bibinfo {pages} {1} (\bibinfo {year} {2020})}\BibitemShut
  {NoStop}%
\bibitem [{\citenamefont {Soundararajan}\ \emph {et~al.}(2021)\citenamefont
  {Soundararajan}, \citenamefont {Sofia}, \citenamefont {Barletta},\ and\
  \citenamefont {Poletto}}]{SoundararajanPodwerBedFusionReviewModelling2021}%
  \BibitemOpen
  \bibfield  {author} {\bibinfo {author} {\bibfnamefont {B.}~\bibnamefont
  {Soundararajan}}, \bibinfo {author} {\bibfnamefont {D.}~\bibnamefont
  {Sofia}}, \bibinfo {author} {\bibfnamefont {D.}~\bibnamefont {Barletta}}, \
  and\ \bibinfo {author} {\bibfnamefont {M.}~\bibnamefont {Poletto}},\ }\href
  {\doibase https://doi.org/10.1016/j.addma.2021.102336} {\bibfield  {journal}
  {\bibinfo  {journal} {Addit. Manuf.}\ }\textbf {\bibinfo {volume} {47}},\
  \bibinfo {pages} {102336} (\bibinfo {year} {2021})}\BibitemShut {NoStop}%
\bibitem [{\citenamefont {Haeri}\ \emph {et~al.}(2017)\citenamefont {Haeri},
  \citenamefont {Wang}, \citenamefont {Ghita},\ and\ \citenamefont
  {Sun}}]{HaeriDEMSpreading2017}%
  \BibitemOpen
  \bibfield  {author} {\bibinfo {author} {\bibfnamefont {S.}~\bibnamefont
  {Haeri}}, \bibinfo {author} {\bibfnamefont {Y.}~\bibnamefont {Wang}},
  \bibinfo {author} {\bibfnamefont {O.}~\bibnamefont {Ghita}}, \ and\ \bibinfo
  {author} {\bibfnamefont {J.}~\bibnamefont {Sun}},\ }\href {\doibase
  https://doi.org/10.1016/j.powtec.2016.11.002} {\bibfield  {journal} {\bibinfo
   {journal} {Powder Technol.}\ }\textbf {\bibinfo {volume} {306}},\ \bibinfo
  {pages} {45} (\bibinfo {year} {2017})}\BibitemShut {NoStop}%
\bibitem [{\citenamefont {Haeri}(2017)}]{HaeriOptimisationBladeSpreaders2017}%
  \BibitemOpen
  \bibfield  {author} {\bibinfo {author} {\bibfnamefont {S.}~\bibnamefont
  {Haeri}},\ }\href {\doibase https://doi.org/10.1016/j.powtec.2017.08.011}
  {\bibfield  {journal} {\bibinfo  {journal} {Powder Technol.}\ }\textbf
  {\bibinfo {volume} {321}},\ \bibinfo {pages} {94} (\bibinfo {year}
  {2017})}\BibitemShut {NoStop}%
\bibitem [{\citenamefont {Ganesan}\ \emph {et~al.}(2008)\citenamefont
  {Ganesan}, \citenamefont {Rosentrater},\ and\ \citenamefont
  {Muthukumarappan}}]{GanesanFlowabilityBulkSolidPowdersReview2008}%
  \BibitemOpen
  \bibfield  {author} {\bibinfo {author} {\bibfnamefont {V.}~\bibnamefont
  {Ganesan}}, \bibinfo {author} {\bibfnamefont {K.}~\bibnamefont
  {Rosentrater}}, \ and\ \bibinfo {author} {\bibfnamefont {K.}~\bibnamefont
  {Muthukumarappan}},\ }\href {\doibase
  https://doi.org/10.1016/j.biosystemseng.2008.09.008} {\bibfield  {journal}
  {\bibinfo  {journal} {Biosyst. Eng.}\ }\textbf {\bibinfo {volume} {101}},\
  \bibinfo {pages} {425} (\bibinfo {year} {2008})}\BibitemShut {NoStop}%
\bibitem [{\citenamefont {Jian}\ \emph {et~al.}(2019)\citenamefont {Jian},
  \citenamefont {Narendran},\ and\ \citenamefont
  {Jayas}}]{JianSegregationBulkGrains2019}%
  \BibitemOpen
  \bibfield  {author} {\bibinfo {author} {\bibfnamefont {F.}~\bibnamefont
  {Jian}}, \bibinfo {author} {\bibfnamefont {R.~B.}\ \bibnamefont {Narendran}},
  \ and\ \bibinfo {author} {\bibfnamefont {D.~S.}\ \bibnamefont {Jayas}},\
  }\href {\doibase https://doi.org/10.1016/j.jspr.2018.12.004} {\bibfield
  {journal} {\bibinfo  {journal} {J. Stored Prod. Res.}\ }\textbf {\bibinfo
  {volume} {81}},\ \bibinfo {pages} {11} (\bibinfo {year} {2019})}\BibitemShut
  {NoStop}%
\bibitem [{\citenamefont {Brennen}(2005)}]{BrennenGranularMultiphaseFlows2005}%
  \BibitemOpen
  \bibfield  {author} {\bibinfo {author} {\bibfnamefont {C.~E.}\ \bibnamefont
  {Brennen}},\ }\enquote {\bibinfo {title} {{Granular Flows}},}\ in\ \href
  {\doibase 10.1017/CBO9780511807169.014} {\emph {\bibinfo {booktitle}
  {Fundamentals of Multiphase Flow}}}\ (\bibinfo  {publisher} {Cambridge
  University Press},\ \bibinfo {year} {2005})\ pp.\ \bibinfo {pages}
  {252--271}\BibitemShut {NoStop}%
\bibitem [{\citenamefont {Forterre}\ and\ \citenamefont
  {Pouliquen}(2008)}]{ForterreOlivierDenseGranularFlows2008}%
  \BibitemOpen
  \bibfield  {author} {\bibinfo {author} {\bibfnamefont {Y.}~\bibnamefont
  {Forterre}}\ and\ \bibinfo {author} {\bibfnamefont {O.}~\bibnamefont
  {Pouliquen}},\ }\href {\doibase 10.1146/annurev.fluid.40.111406.102142}
  {\bibfield  {journal} {\bibinfo  {journal} {Annu. Rev. Fluid Mech.}\ }\textbf
  {\bibinfo {volume} {40}},\ \bibinfo {pages} {1} (\bibinfo {year}
  {2008})}\BibitemShut {NoStop}%
\bibitem [{\citenamefont {Rao}(2008)}]{RaoGranularFlowIntro2008}%
  \BibitemOpen
  \bibfield  {author} {\bibinfo {author} {\bibfnamefont {P.~R.}\ \bibnamefont
  {Rao}, \bibfnamefont {K.~K.and~Nott}},\ }\href@noop {} {\emph {\bibinfo
  {title} {{An Introduction to Granular Flow}}}}\ (\bibinfo  {publisher}
  {Cambridge University Press, New York},\ \bibinfo {year} {2008})\BibitemShut
  {NoStop}%
\bibitem [{\citenamefont {{da Cruz}}\ \emph {et~al.}(2003)\citenamefont {{da
  Cruz}}, \citenamefont {Chevoir}, \citenamefont {Roux},\ and\ \citenamefont
  {Iordanoff}}]{daCruzMacroscopicFrictionGranular2003}%
  \BibitemOpen
  \bibfield  {author} {\bibinfo {author} {\bibfnamefont {F.}~\bibnamefont {{da
  Cruz}}}, \bibinfo {author} {\bibfnamefont {F.}~\bibnamefont {Chevoir}},
  \bibinfo {author} {\bibfnamefont {J.-N.}\ \bibnamefont {Roux}}, \ and\
  \bibinfo {author} {\bibfnamefont {I.}~\bibnamefont {Iordanoff}},\ }in\ \href
  {\doibase https://doi.org/10.1016/S0167-8922(03)80034-2} {\emph {\bibinfo
  {booktitle} {Transient Processes in Tribology}}},\ \bibinfo {series}
  {Tribology Series}, Vol.~\bibinfo {volume} {43},\ \bibinfo {editor} {edited
  by\ \bibinfo {editor} {\bibfnamefont {G.}~\bibnamefont {Dalmaz}}, \bibinfo
  {editor} {\bibfnamefont {A.}~\bibnamefont {Lubrecht}}, \bibinfo {editor}
  {\bibfnamefont {D.}~\bibnamefont {Dowson}}, \ and\ \bibinfo {editor}
  {\bibfnamefont {M.}~\bibnamefont {Priest}}}\ (\bibinfo  {publisher}
  {Elsevier},\ \bibinfo {year} {2003})\ pp.\ \bibinfo {pages}
  {53--61}\BibitemShut {NoStop}%
\bibitem [{\citenamefont {Sela}\ and\ \citenamefont
  {Goldhirsch}(1998)}]{SG1998}%
  \BibitemOpen
  \bibfield  {author} {\bibinfo {author} {\bibfnamefont {N.}~\bibnamefont
  {Sela}}\ and\ \bibinfo {author} {\bibfnamefont {I.}~\bibnamefont
  {Goldhirsch}},\ }\href@noop {} {\bibfield  {journal} {\bibinfo  {journal}
  {J.~Fluid Mech.}\ }\textbf {\bibinfo {volume} {361}},\ \bibinfo {pages} {41}
  (\bibinfo {year} {1998})}\BibitemShut {NoStop}%
\bibitem [{\citenamefont {Garz\'o}\ and\ \citenamefont
  {Dufty}(1999)}]{GarzoDuftyGranularKinetic1999}%
  \BibitemOpen
  \bibfield  {author} {\bibinfo {author} {\bibfnamefont {V.}~\bibnamefont
  {Garz\'o}}\ and\ \bibinfo {author} {\bibfnamefont {J.~W.}\ \bibnamefont
  {Dufty}},\ }\href {\doibase 10.1103/PhysRevE.59.5895} {\bibfield  {journal}
  {\bibinfo  {journal} {Phys. Rev. E}\ }\textbf {\bibinfo {volume} {59}},\
  \bibinfo {pages} {5895} (\bibinfo {year} {1999})}\BibitemShut {NoStop}%
\bibitem [{\citenamefont
  {Santos}(2003)}]{SantosTraansportCoefficientsInelasticMaxwell2003}%
  \BibitemOpen
  \bibfield  {author} {\bibinfo {author} {\bibfnamefont {A.}~\bibnamefont
  {Santos}},\ }\href {\doibase https://doi.org/10.1016/S0378-4371(02)01005-1}
  {\bibfield  {journal} {\bibinfo  {journal} {Phys. A (Amsterdam, Neth.)}\
  }\textbf {\bibinfo {volume} {321}},\ \bibinfo {pages} {442} (\bibinfo {year}
  {2003})}\BibitemShut {NoStop}%
\bibitem [{\citenamefont {Nott}(2011)}]{NottBCWallGranular2011}%
  \BibitemOpen
  \bibfield  {author} {\bibinfo {author} {\bibfnamefont {P.~R.}\ \bibnamefont
  {Nott}},\ }\href {\doibase 10.1017/jfm.2011.105} {\bibfield  {journal}
  {\bibinfo  {journal} {J. Fluid Mech.}\ }\textbf {\bibinfo {volume} {678}},\
  \bibinfo {pages} {179} (\bibinfo {year} {2011})}\BibitemShut {NoStop}%
\bibitem [{\citenamefont {Lun}\ \emph {et~al.}(1984)\citenamefont {Lun},
  \citenamefont {Savage}, \citenamefont {Jeffrey},\ and\ \citenamefont
  {Chepurniy}}]{LunKineticGranularCouette1984}%
  \BibitemOpen
  \bibfield  {author} {\bibinfo {author} {\bibfnamefont {C.~K.~K.}\
  \bibnamefont {Lun}}, \bibinfo {author} {\bibfnamefont {S.~B.}\ \bibnamefont
  {Savage}}, \bibinfo {author} {\bibfnamefont {D.~J.}\ \bibnamefont {Jeffrey}},
  \ and\ \bibinfo {author} {\bibfnamefont {N.}~\bibnamefont {Chepurniy}},\
  }\href {\doibase 10.1017/S0022112084000586} {\bibfield  {journal} {\bibinfo
  {journal} {J. Fluid Mech.}\ }\textbf {\bibinfo {volume} {140}},\ \bibinfo
  {pages} {223} (\bibinfo {year} {1984})}\BibitemShut {NoStop}%
\bibitem [{\citenamefont {Abu‐Zaid}\ and\ \citenamefont
  {Ahmadi}(1990)}]{AbuZaidKineticGranularFrictional1990}%
  \BibitemOpen
  \bibfield  {author} {\bibinfo {author} {\bibfnamefont {S.}~\bibnamefont
  {Abu‐Zaid}}\ and\ \bibinfo {author} {\bibfnamefont {G.}~\bibnamefont
  {Ahmadi}},\ }\href {\doibase 10.1061/(ASCE)0733-9399(1990)116:2(379)}
  {\bibfield  {journal} {\bibinfo  {journal} {ASCE J. Eng. Mech. Div.}\
  }\textbf {\bibinfo {volume} {116}},\ \bibinfo {pages} {379} (\bibinfo {year}
  {1990})}\BibitemShut {NoStop}%
\bibitem [{\citenamefont {Lun}\ and\ \citenamefont
  {Savage}(1987)}]{LunKineticGranularRoughInelastic1987}%
  \BibitemOpen
  \bibfield  {author} {\bibinfo {author} {\bibfnamefont {C.~K.~K.}\
  \bibnamefont {Lun}}\ and\ \bibinfo {author} {\bibfnamefont {S.~B.}\
  \bibnamefont {Savage}},\ }\href {\doibase 10.1115/1.3172993} {\bibfield
  {journal} {\bibinfo  {journal} {J. Appl. Mech.}\ }\textbf {\bibinfo {volume}
  {54}},\ \bibinfo {pages} {47} (\bibinfo {year} {1987})}\BibitemShut {NoStop}%
\bibitem [{\citenamefont {Chialvo}\ and\ \citenamefont
  {Sundaresan}(2013)}]{ChialvoKineticFrictionGranular2013}%
  \BibitemOpen
  \bibfield  {author} {\bibinfo {author} {\bibfnamefont {S.}~\bibnamefont
  {Chialvo}}\ and\ \bibinfo {author} {\bibfnamefont {S.}~\bibnamefont
  {Sundaresan}},\ }\href {\doibase 10.1063/1.4812804} {\bibfield  {journal}
  {\bibinfo  {journal} {Phys. Fluids}\ }\textbf {\bibinfo {volume} {25}},\
  \bibinfo {pages} {070603} (\bibinfo {year} {2013})}\BibitemShut {NoStop}%
\bibitem [{\citenamefont {Vescovi}\ \emph {et~al.}(2014)\citenamefont
  {Vescovi}, \citenamefont {Berzi}, \citenamefont {Richard},\ and\
  \citenamefont {Brodu}}]{VescoviGranularShearCouetteDEM2014}%
  \BibitemOpen
  \bibfield  {author} {\bibinfo {author} {\bibfnamefont {D.}~\bibnamefont
  {Vescovi}}, \bibinfo {author} {\bibfnamefont {D.}~\bibnamefont {Berzi}},
  \bibinfo {author} {\bibfnamefont {P.}~\bibnamefont {Richard}}, \ and\
  \bibinfo {author} {\bibfnamefont {N.}~\bibnamefont {Brodu}},\ }\href
  {\doibase 10.1063/1.4879267} {\bibfield  {journal} {\bibinfo  {journal}
  {Phys. Fluids}\ }\textbf {\bibinfo {volume} {26}},\ \bibinfo {pages} {053305}
  (\bibinfo {year} {2014})}\BibitemShut {NoStop}%
\bibitem [{\citenamefont {Campbell}\ and\ \citenamefont
  {Brennen}(1985)}]{CampbellBrennenSimulationGranular1985}%
  \BibitemOpen
  \bibfield  {author} {\bibinfo {author} {\bibfnamefont {C.~S.}\ \bibnamefont
  {Campbell}}\ and\ \bibinfo {author} {\bibfnamefont {C.~E.}\ \bibnamefont
  {Brennen}},\ }\href {\doibase 10.1017/S002211208500091X} {\bibfield
  {journal} {\bibinfo  {journal} {J. Fluid Mech.}\ }\textbf {\bibinfo {volume}
  {151}},\ \bibinfo {pages} {167} (\bibinfo {year} {1985})}\BibitemShut
  {NoStop}%
\bibitem [{\citenamefont {Richman}(1988)}]{RichmanBC1988}%
  \BibitemOpen
  \bibfield  {author} {\bibinfo {author} {\bibfnamefont {M.~W.}\ \bibnamefont
  {Richman}},\ }\href {\doibase 10.1007/BF01174637} {\bibfield  {journal}
  {\bibinfo  {journal} {Acta Mech.}\ }\textbf {\bibinfo {volume} {75}},\
  \bibinfo {pages} {227} (\bibinfo {year} {1988})}\BibitemShut {NoStop}%
\bibitem [{\citenamefont
  {Dadzie}(2013)}]{DadzieThermomechanicalContinuumFlow2013}%
  \BibitemOpen
  \bibfield  {author} {\bibinfo {author} {\bibfnamefont {S.~K.}\ \bibnamefont
  {Dadzie}},\ }\href {\doibase 10.1017/jfm.2012.546} {\bibfield  {journal}
  {\bibinfo  {journal} {J. Fluid Mech.}\ }\textbf {\bibinfo {volume} {716}},\
  \bibinfo {pages} {R6} (\bibinfo {year} {2013})}\BibitemShut {NoStop}%
\bibitem [{\citenamefont {Wu}\ \emph {et~al.}(2016)\citenamefont {Wu},
  \citenamefont {Liu}, \citenamefont {Reese},\ and\ \citenamefont
  {Zhang}}]{WuGranularGasPoiseuille2016}%
  \BibitemOpen
  \bibfield  {author} {\bibinfo {author} {\bibfnamefont {L.}~\bibnamefont
  {Wu}}, \bibinfo {author} {\bibfnamefont {H.}~\bibnamefont {Liu}}, \bibinfo
  {author} {\bibfnamefont {J.~M.}\ \bibnamefont {Reese}}, \ and\ \bibinfo
  {author} {\bibfnamefont {Y.}~\bibnamefont {Zhang}},\ }\href {\doibase
  10.1017/jfm.2016.173} {\bibfield  {journal} {\bibinfo  {journal} {J. Fluid
  Mech.}\ }\textbf {\bibinfo {volume} {794}},\ \bibinfo {pages} {252} (\bibinfo
  {year} {2016})}\BibitemShut {NoStop}%
\bibitem [{\citenamefont {Dadzie}\ and\ \citenamefont
  {Christou}(2016)}]{DadzieBivelocityGasLidCavity2016}%
  \BibitemOpen
  \bibfield  {author} {\bibinfo {author} {\bibfnamefont {S.~K.}\ \bibnamefont
  {Dadzie}}\ and\ \bibinfo {author} {\bibfnamefont {C.}~\bibnamefont
  {Christou}},\ }\href {\doibase
  https://doi.org/10.1016/j.icheatmasstransfer.2016.09.006} {\bibfield
  {journal} {\bibinfo  {journal} {Int. Commun. Heat Mass Transfer}\ }\textbf
  {\bibinfo {volume} {78}},\ \bibinfo {pages} {175} (\bibinfo {year}
  {2016})}\BibitemShut {NoStop}%
\bibitem [{\citenamefont {Christou}\ and\ \citenamefont
  {Dadzie}(2017)}]{ChristouHeatTransferCavity2017}%
  \BibitemOpen
  \bibfield  {author} {\bibinfo {author} {\bibfnamefont {C.}~\bibnamefont
  {Christou}}\ and\ \bibinfo {author} {\bibfnamefont {S.~K.}\ \bibnamefont
  {Dadzie}},\ }\href {\doibase 10.1115/1.4036340} {\bibfield  {journal}
  {\bibinfo  {journal} {J. Heat Transfer}\ }\textbf {\bibinfo {volume} {139}}
  (\bibinfo {year} {2017}),\ 10.1115/1.4036340},\ \bibinfo {note}
  {092002}\BibitemShut {NoStop}%
\bibitem [{\citenamefont {Christou}\ and\ \citenamefont
  {Dadzie}(2018)}]{ChristouRarefiedGasRecasted2018}%
  \BibitemOpen
  \bibfield  {author} {\bibinfo {author} {\bibfnamefont {C.}~\bibnamefont
  {Christou}}\ and\ \bibinfo {author} {\bibfnamefont {S.~K.}\ \bibnamefont
  {Dadzie}},\ }\href {\doibase 10.1088/2399-6528/aab066} {\bibfield  {journal}
  {\bibinfo  {journal} {J. Phys. Commun.}\ }\textbf {\bibinfo {volume} {2}},\
  \bibinfo {pages} {035002} (\bibinfo {year} {2018})}\BibitemShut {NoStop}%
\bibitem [{\citenamefont {Stamatiou}\ \emph {et~al.}(2019)\citenamefont
  {Stamatiou}, \citenamefont {Dadzie},\ and\ \citenamefont
  {Reddy}}]{StamatiouEnhancedFlowNanotube2019}%
  \BibitemOpen
  \bibfield  {author} {\bibinfo {author} {\bibfnamefont {A.}~\bibnamefont
  {Stamatiou}}, \bibinfo {author} {\bibfnamefont {S.~K.}\ \bibnamefont
  {Dadzie}}, \ and\ \bibinfo {author} {\bibfnamefont {M.~H.~L.}\ \bibnamefont
  {Reddy}},\ }\href {\doibase 10.1088/2399-6528/ab5f9e} {\bibfield  {journal}
  {\bibinfo  {journal} {J. Phys. Commun.}\ }\textbf {\bibinfo {volume} {3}},\
  \bibinfo {pages} {125012} (\bibinfo {year} {2019})}\BibitemShut {NoStop}%
\bibitem [{\citenamefont {Lockerby}\ and\ \citenamefont
  {Reese}(2003)}]{LockerbyBurnettMicroCouette2003}%
  \BibitemOpen
  \bibfield  {author} {\bibinfo {author} {\bibfnamefont {D.~A.}\ \bibnamefont
  {Lockerby}}\ and\ \bibinfo {author} {\bibfnamefont {J.~M.}\ \bibnamefont
  {Reese}},\ }\href {\doibase https://doi.org/10.1016/S0021-9991(03)00162-1}
  {\bibfield  {journal} {\bibinfo  {journal} {J. Comp. Phys.}\ }\textbf
  {\bibinfo {volume} {188}},\ \bibinfo {pages} {333} (\bibinfo {year}
  {2003})}\BibitemShut {NoStop}%
\bibitem [{\citenamefont {Reddy}\ and\ \citenamefont {Alam}(2020)}]{Reddy2020}%
  \BibitemOpen
  \bibfield  {author} {\bibinfo {author} {\bibfnamefont {M.~H.~L.}\
  \bibnamefont {Reddy}}\ and\ \bibinfo {author} {\bibfnamefont
  {M.}~\bibnamefont {Alam}},\ }\href {\doibase 10.1103/PhysRevFluids.5.044302}
  {\bibfield  {journal} {\bibinfo  {journal} {Phys. Rev. Fluids}\ }\textbf
  {\bibinfo {volume} {5}},\ \bibinfo {pages} {044302} (\bibinfo {year}
  {2020})}\BibitemShut {NoStop}%
\bibitem [{\citenamefont {Knudsen}(1909)}]{KnudsenMinimum1909}%
  \BibitemOpen
  \bibfield  {author} {\bibinfo {author} {\bibfnamefont {M.}~\bibnamefont
  {Knudsen}},\ }\href {\doibase https://doi.org/10.1002/andp.19093330106}
  {\bibfield  {journal} {\bibinfo  {journal} {Ann. Phys. (Berlin, Ger.)}\
  }\textbf {\bibinfo {volume} {333}},\ \bibinfo {pages} {75} (\bibinfo {year}
  {1909})}\BibitemShut {NoStop}%
\bibitem [{\citenamefont {Dadzie}\ and\ \citenamefont
  {Brenner}(2012)}]{DadzieEnhancedMicroChannels2012}%
  \BibitemOpen
  \bibfield  {author} {\bibinfo {author} {\bibfnamefont {S.~K.}\ \bibnamefont
  {Dadzie}}\ and\ \bibinfo {author} {\bibfnamefont {H.}~\bibnamefont
  {Brenner}},\ }\href {\doibase 10.1103/PhysRevE.86.036318} {\bibfield
  {journal} {\bibinfo  {journal} {Phys. Rev. E}\ }\textbf {\bibinfo {volume}
  {86}},\ \bibinfo {pages} {036318} (\bibinfo {year} {2012})}\BibitemShut
  {NoStop}%
\bibitem [{\citenamefont
  {Brenner}(2012{\natexlab{a}})}]{BrennerBivelocity2012}%
  \BibitemOpen
  \bibfield  {author} {\bibinfo {author} {\bibfnamefont {H.}~\bibnamefont
  {Brenner}},\ }\href {\doibase https://doi.org/10.1016/j.ijengsci.2012.01.006}
  {\bibfield  {journal} {\bibinfo  {journal} {Int. J. Eng. Sci.}\ }\textbf
  {\bibinfo {volume} {54}},\ \bibinfo {pages} {67} (\bibinfo {year}
  {2012}{\natexlab{a}})}\BibitemShut {NoStop}%
\bibitem [{\citenamefont
  {Brenner}(2012{\natexlab{b}})}]{BrennerFluidMechanicsRestVD2012}%
  \BibitemOpen
  \bibfield  {author} {\bibinfo {author} {\bibfnamefont {H.}~\bibnamefont
  {Brenner}},\ }\href {\doibase 10.1103/PhysRevE.86.016307} {\bibfield
  {journal} {\bibinfo  {journal} {Phys. Rev. E}\ }\textbf {\bibinfo {volume}
  {86}},\ \bibinfo {pages} {016307} (\bibinfo {year}
  {2012}{\natexlab{b}})}\BibitemShut {NoStop}%
\bibitem [{\citenamefont {Reddy}\ and\ \citenamefont
  {Dadzie}(2021)}]{LakshminarayanaMolecuarDiffusivityShockWave2021}%
  \BibitemOpen
  \bibfield  {author} {\bibinfo {author} {\bibfnamefont {L.~M.~H.}\
  \bibnamefont {Reddy}}\ and\ \bibinfo {author} {\bibfnamefont {S.~K.}\
  \bibnamefont {Dadzie}},\ }\href {\doibase 10.1103/PhysRevE.104.035111}
  {\bibfield  {journal} {\bibinfo  {journal} {Phys. Rev. E}\ }\textbf {\bibinfo
  {volume} {104}},\ \bibinfo {pages} {035111} (\bibinfo {year}
  {2021})}\BibitemShut {NoStop}%
\bibitem [{\citenamefont {Kloss}\ \emph {et~al.}(2012)\citenamefont {Kloss},
  \citenamefont {Goniva}, \citenamefont {Hager}, \citenamefont {Amberger},\
  and\ \citenamefont {Pirker}}]{KlossLIGGGHTS2012}%
  \BibitemOpen
  \bibfield  {author} {\bibinfo {author} {\bibfnamefont {C.}~\bibnamefont
  {Kloss}}, \bibinfo {author} {\bibfnamefont {C.}~\bibnamefont {Goniva}},
  \bibinfo {author} {\bibfnamefont {A.}~\bibnamefont {Hager}}, \bibinfo
  {author} {\bibfnamefont {S.}~\bibnamefont {Amberger}}, \ and\ \bibinfo
  {author} {\bibfnamefont {S.}~\bibnamefont {Pirker}},\ }\href {\doibase
  10.1504/PCFD.2012.047457} {\bibfield  {journal} {\bibinfo  {journal} {Prog.
  Comput. Fluid Dyn.}\ }\textbf {\bibinfo {volume} {12}},\ \bibinfo {pages}
  {140} (\bibinfo {year} {2012})}\BibitemShut {NoStop}%
\bibitem [{\citenamefont {Nott}\ \emph {et~al.}(1999)\citenamefont {Nott},
  \citenamefont {Alam}, \citenamefont {Agrawal}, \citenamefont {Jackson},\ and\
  \citenamefont {Sundaresan}}]{NottBoundariesCouetteBifurcation1999}%
  \BibitemOpen
  \bibfield  {author} {\bibinfo {author} {\bibfnamefont {P.~R.}\ \bibnamefont
  {Nott}}, \bibinfo {author} {\bibfnamefont {M.}~\bibnamefont {Alam}}, \bibinfo
  {author} {\bibfnamefont {K.}~\bibnamefont {Agrawal}}, \bibinfo {author}
  {\bibfnamefont {R.}~\bibnamefont {Jackson}}, \ and\ \bibinfo {author}
  {\bibfnamefont {S.}~\bibnamefont {Sundaresan}},\ }\href {\doibase
  10.1017/S0022112099006060} {\bibfield  {journal} {\bibinfo  {journal} {J.
  Fluid Mech.}\ }\textbf {\bibinfo {volume} {397}},\ \bibinfo {pages} {203}
  (\bibinfo {year} {1999})}\BibitemShut {NoStop}%
\bibitem [{\citenamefont {Alam}\ and\ \citenamefont
  {Nott}(1998)}]{AlamStabilityCouetteFlow1998}%
  \BibitemOpen
  \bibfield  {author} {\bibinfo {author} {\bibfnamefont {M.}~\bibnamefont
  {Alam}}\ and\ \bibinfo {author} {\bibfnamefont {P.~R.}\ \bibnamefont
  {Nott}},\ }\href {\doibase 10.1017/S002211209800295X} {\bibfield  {journal}
  {\bibinfo  {journal} {J. Fluid Mech.}\ }\textbf {\bibinfo {volume} {377}},\
  \bibinfo {pages} {99} (\bibinfo {year} {1998})}\BibitemShut {NoStop}%
\bibitem [{\citenamefont {Alam}\ \emph {et~al.}(2005)\citenamefont {Alam},
  \citenamefont {Arakeri}, \citenamefont {Nott}, \citenamefont {Goddard},\ and\
  \citenamefont {Hermann}}]{AlamInstabilityOrderingGravity2005}%
  \BibitemOpen
  \bibfield  {author} {\bibinfo {author} {\bibfnamefont {M.}~\bibnamefont
  {Alam}}, \bibinfo {author} {\bibfnamefont {V.~H.}\ \bibnamefont {Arakeri}},
  \bibinfo {author} {\bibfnamefont {P.~R.}\ \bibnamefont {Nott}}, \bibinfo
  {author} {\bibfnamefont {J.~D.}\ \bibnamefont {Goddard}}, \ and\ \bibinfo
  {author} {\bibfnamefont {H.~J.}\ \bibnamefont {Hermann}},\ }\href {\doibase
  10.1017/S0022112004002150} {\bibfield  {journal} {\bibinfo  {journal} {J.
  Fluid Mech.}\ }\textbf {\bibinfo {volume} {523}},\ \bibinfo {pages} {277}
  (\bibinfo {year} {2005})}\BibitemShut {NoStop}%
\bibitem [{\citenamefont {Carnahan}\ and\ \citenamefont
  {Starling}(1969)}]{CarnahnStarlingg01969}%
  \BibitemOpen
  \bibfield  {author} {\bibinfo {author} {\bibfnamefont {N.~F.}\ \bibnamefont
  {Carnahan}}\ and\ \bibinfo {author} {\bibfnamefont {K.~E.}\ \bibnamefont
  {Starling}},\ }\href {\doibase 10.1063/1.1672048} {\bibfield  {journal}
  {\bibinfo  {journal} {J. Chem. Phys.}\ }\textbf {\bibinfo {volume} {51}},\
  \bibinfo {pages} {635} (\bibinfo {year} {1969})}\BibitemShut {NoStop}%
\bibitem [{\citenamefont {Dadzie}\ \emph {et~al.}(2008)\citenamefont {Dadzie},
  \citenamefont {Reese},\ and\ \citenamefont
  {McInnes}}]{DadzieContinuumGasDensityVariation2008}%
  \BibitemOpen
  \bibfield  {author} {\bibinfo {author} {\bibfnamefont {S.~K.}\ \bibnamefont
  {Dadzie}}, \bibinfo {author} {\bibfnamefont {J.~M.}\ \bibnamefont {Reese}}, \
  and\ \bibinfo {author} {\bibfnamefont {C.~R.}\ \bibnamefont {McInnes}},\
  }\href {\doibase https://doi.org/10.1016/j.physa.2008.07.009} {\bibfield
  {journal} {\bibinfo  {journal} {Phys. A (Amsterdam, Neth.)}\ }\textbf
  {\bibinfo {volume} {387}},\ \bibinfo {pages} {6079} (\bibinfo {year}
  {2008})}\BibitemShut {NoStop}%
\bibitem [{\citenamefont {Berzi}(2014)}]{BerziExtendedKineticTheory2014}%
  \BibitemOpen
  \bibfield  {author} {\bibinfo {author} {\bibfnamefont {D.}~\bibnamefont
  {Berzi}},\ }\href {\doibase 10.1007/s00707-014-1125-1} {\bibfield  {journal}
  {\bibinfo  {journal} {Acta Mech.}\ }\textbf {\bibinfo {volume} {225}},\
  \bibinfo {pages} {2191} (\bibinfo {year} {2014})}\BibitemShut {NoStop}%
\bibitem [{\citenamefont {Jenkins}\ and\ \citenamefont
  {Berzi}(2012)}]{JenkinsBerziKineticTheoryGranularInclined2012}%
  \BibitemOpen
  \bibfield  {author} {\bibinfo {author} {\bibfnamefont {J.~T.}\ \bibnamefont
  {Jenkins}}\ and\ \bibinfo {author} {\bibfnamefont {D.}~\bibnamefont
  {Berzi}},\ }\href {\doibase 10.1007/s10035-011-0308-x} {\bibfield  {journal}
  {\bibinfo  {journal} {Granul. Matter}\ }\textbf {\bibinfo {volume} {14}},\
  \bibinfo {pages} {79} (\bibinfo {year} {2012})}\BibitemShut {NoStop}%
\bibitem [{\citenamefont {Tomy}\ and\ \citenamefont
  {Dadzie}(2022)}]{TomyDiffusionSlip2022}%
  \BibitemOpen
  \bibfield  {author} {\bibinfo {author} {\bibfnamefont {A.~M.}\ \bibnamefont
  {Tomy}}\ and\ \bibinfo {author} {\bibfnamefont {S.~K.}\ \bibnamefont
  {Dadzie}},\ }\href {\doibase 10.3390/mi13091425} {\bibfield  {journal}
  {\bibinfo  {journal} {Micromachines}\ }\textbf {\bibinfo {volume} {13}}
  (\bibinfo {year} {2022}),\ 10.3390/mi13091425}\BibitemShut {NoStop}%
\bibitem [{\citenamefont {Reddy}\ \emph {et~al.}(2019)\citenamefont {Reddy},
  \citenamefont {Dadzie}, \citenamefont {Ocone}, \citenamefont {Borg},\ and\
  \citenamefont {Reese}}]{LakshminarayanaRNS2019}%
  \BibitemOpen
  \bibfield  {author} {\bibinfo {author} {\bibfnamefont {M.~H.~L.}\
  \bibnamefont {Reddy}}, \bibinfo {author} {\bibfnamefont {S.~K.}\ \bibnamefont
  {Dadzie}}, \bibinfo {author} {\bibfnamefont {R.}~\bibnamefont {Ocone}},
  \bibinfo {author} {\bibfnamefont {M.~K.}\ \bibnamefont {Borg}}, \ and\
  \bibinfo {author} {\bibfnamefont {J.~M.}\ \bibnamefont {Reese}},\ }\href
  {\doibase 10.1088/2399-6528/ab4b86} {\bibfield  {journal} {\bibinfo
  {journal} {J. Phys. Commun.}\ }\textbf {\bibinfo {volume} {3}},\ \bibinfo
  {pages} {105009} (\bibinfo {year} {2019})}\BibitemShut {NoStop}%
\bibitem [{\citenamefont {Dadzie}\ and\ \citenamefont
  {Reese}(2012{\natexlab{a}})}]{DadzieReeseThermomechanicalHydrodynamicKnudsen2012}%
  \BibitemOpen
  \bibfield  {author} {\bibinfo {author} {\bibfnamefont {S.~K.}\ \bibnamefont
  {Dadzie}}\ and\ \bibinfo {author} {\bibfnamefont {J.~M.}\ \bibnamefont
  {Reese}},\ }\href {\doibase 10.1103/PhysRevE.85.041202} {\bibfield  {journal}
  {\bibinfo  {journal} {Phys. Rev. E}\ }\textbf {\bibinfo {volume} {85}},\
  \bibinfo {pages} {041202} (\bibinfo {year} {2012}{\natexlab{a}})}\BibitemShut
  {NoStop}%
\bibitem [{\citenamefont {Dadzie}\ and\ \citenamefont
  {Reese}(2012{\natexlab{b}})}]{DadzieReeseSpatialStochasityGasFlows2012}%
  \BibitemOpen
  \bibfield  {author} {\bibinfo {author} {\bibfnamefont {S.~K.}\ \bibnamefont
  {Dadzie}}\ and\ \bibinfo {author} {\bibfnamefont {J.~M.}\ \bibnamefont
  {Reese}},\ }\href {\doibase https://doi.org/10.1016/j.physleta.2012.01.004}
  {\bibfield  {journal} {\bibinfo  {journal} {Phys. Lett. A}\ }\textbf
  {\bibinfo {volume} {376}},\ \bibinfo {pages} {967} (\bibinfo {year}
  {2012}{\natexlab{b}})}\BibitemShut {NoStop}%
\bibitem [{\citenamefont {Berloff}\ \emph {et~al.}(2014)\citenamefont
  {Berloff}, \citenamefont {Brachet},\ and\ \citenamefont
  {Proukakis}}]{BerloffFluidMechanicaNonZerpTemp2014}%
  \BibitemOpen
  \bibfield  {author} {\bibinfo {author} {\bibfnamefont {N.~G.}\ \bibnamefont
  {Berloff}}, \bibinfo {author} {\bibfnamefont {M.}~\bibnamefont {Brachet}}, \
  and\ \bibinfo {author} {\bibfnamefont {N.~P.}\ \bibnamefont {Proukakis}},\
  }\href {\doibase 10.1073/pnas.1312549111} {\bibfield  {journal} {\bibinfo
  {journal} {Proc. Natl. Acad. Sci. U. S. A.}\ }\textbf {\bibinfo {volume}
  {111}},\ \bibinfo {pages} {4675} (\bibinfo {year} {2014})}\BibitemShut
  {NoStop}%
\bibitem [{\citenamefont {Reddy}\ and\ \citenamefont {Alam}(2015)}]{Reddy2015}%
  \BibitemOpen
  \bibfield  {author} {\bibinfo {author} {\bibfnamefont {M.~H.~L.}\
  \bibnamefont {Reddy}}\ and\ \bibinfo {author} {\bibfnamefont
  {M.}~\bibnamefont {Alam}},\ }\href {\doibase 10.1017/jfm.2015.455} {\bibfield
   {journal} {\bibinfo  {journal} {J.~Fluid Mech.}\ }\textbf {\bibinfo {volume}
  {779}},\ \bibinfo {pages} {R2} (\bibinfo {year} {2015})}\BibitemShut
  {NoStop}%
\bibitem [{\citenamefont {Lees}\ and\ \citenamefont
  {Edwards}(1972)}]{LeesEdwardsBC1972}%
  \BibitemOpen
  \bibfield  {author} {\bibinfo {author} {\bibfnamefont {A.~W.}\ \bibnamefont
  {Lees}}\ and\ \bibinfo {author} {\bibfnamefont {S.~F.}\ \bibnamefont
  {Edwards}},\ }\href {\doibase 10.1088/0022-3719/5/15/006} {\bibfield
  {journal} {\bibinfo  {journal} {J. Phys. C: Solid State Phys.}\ }\textbf
  {\bibinfo {volume} {5}},\ \bibinfo {pages} {1921} (\bibinfo {year}
  {1972})}\BibitemShut {NoStop}%
\bibitem [{\citenamefont {MATLAB}(2021)}]{MATLAB:R2021a_u1}%
  \BibitemOpen
  \bibfield  {author} {\bibinfo {author} {\bibnamefont {MATLAB}},\ }\href@noop
  {} {\emph {\bibinfo {title} {{MATLAB version 9.10.0 (R2021a)}}}}\ (\bibinfo
  {publisher} {The Mathworks, Inc.},\ \bibinfo {address} {Natick,
  Massachusetts},\ \bibinfo {year} {2021})\BibitemShut {NoStop}%
\end{thebibliography}

%

\end{document}